\documentclass[aps,prd,floatfix,twocolumn,superscriptaddress,preprintnumbers,nofootinbib]{revtex4-1}
\usepackage{amsmath,amsfonts,amssymb}
\usepackage{graphicx}
\usepackage{epsfig}
\usepackage{color}
\usepackage[dvipsnames,svgnames,x11names]{xcolor}
\usepackage{slashed} % slash mark
\usepackage{url}
\usepackage{subfigure}
\usepackage{multirow} % multirows in tabular environment
\usepackage{comment}
\newcommand{\Dfbd}{\mathord{\buildrel{\lower3pt\hbox{$\scriptscriptstyle\leftrightarrow$}}\over {D}_{\mu}}}
\newcommand{\CO}{\mathcal{O}}
\newcommand{\be}{\begin{equation}}
\newcommand{\ee}{\end{equation}}
\newcommand{\besp}{\begin{equation}\begin{split}}
\newcommand{\eesp}{\end{split}\end{equation}}

\newcommand{\nn}{\nonumber}

\usepackage{ulem}
\usepackage[colorlinks,citecolor=red]{hyperref}
\allowdisplaybreaks[2]

\makeatother

\begin{document}

\title{Testing the electroweak phase transition in scalar extension models at lepton colliders}
\author{Qing-Hong Cao}
\email{qinghongcao@pku.edu.cn}
\affiliation{Department of Physics and State Key Laboratory of Nuclear Physics and Technology,\\ Peking University, Beijing 100871, China}
\affiliation{Collaborative Innovation Center of Quantum Matter, Beijing 100871, China}
\affiliation{Center for High Energy Physics, Peking University, Beijing 100871, China}

\author{Fa Peng Huang}
\email{huangfp@ihep.ac.cn}
\affiliation{Theoretical Physics Division, Institute of High Energy Physics, Chinese Academy of Sciences, P.O.Box 918-4, Beijing 100049, P.R.China}

\author{Ke-Pan Xie}
\email{kpxie@pku.edu.cn}
\affiliation{Department of Physics and State Key Laboratory of Nuclear Physics and Technology,\\ Peking University, Beijing 100871, China}

\author{Xinmin Zhang}
\email{xmzhang@ihep.ac.cn}
\affiliation{Theoretical Physics Division, Institute of High Energy Physics, Chinese Academy of Sciences, P.O.Box 918-4, Beijing 100049, P.R.China}
\affiliation{School of Physics Sciences, University of Chinese Academy of Sciences, Beijing 100039, China}

\begin{abstract}

We study the electroweak phase transition in three scalar extension models beyond the Standard Model. Assuming new scalars are decoupled at some heavy scale, we use the covariant derivative expansion method to derive all of the dimension-6 effective operators, whose coefficients are highly correlated in a specific model. We provide bounds to the complete set of dimension-6 operators by including the electroweak precision test and recent Higgs measurements. We find that the parameter space of strong first-order phase transitions (induced by the $|H|^6$ operator) can be probed extensively  in $Zh$ production at future electron-positron colliders.

\end{abstract}

%\pacs{98.80.Cq,12.60.-i}
\maketitle

\section{Introduction}

After the discovery of the Higgs boson~\cite{Aad:2012tfa,Chatrchyan:2012xdj} it is important to probe the shape of the Higgs potential, which determines the electroweak (EW) symmetry breaking pattern or the type of the EW phase transition.
Current experimental data only tell us local information about the Higgs potential around the vacuum
expectation value $v=246$ GeV, and say nothing about its global features.
Unravelling the properties of the Higgs potential can also shed light on the origin of the baryon asymmetry of the universe.
The baryon asymmetry of the universe is described by the baryon-to-photon ratio
$\eta_B = 6.05(7)\times 10 ^{-10}$~\cite{Ade:2013zuv,Olive:2016xmw}, extracted from
the experimental data on the cosmic microwave background radiation and big bang nucleosynthesis.
Various baryogenesis mechanisms~\cite{Dine:2003ax} have been proposed to satisfy the three Sakharov conditions~\cite{Sakharov:1967dj}.
Among the various mechanisms, EW baryogenesis~\cite{Kuzmin:1985mm,Trodden:1998ym,Morrissey:2012db} becomes a popular and  promising mechanism after the first detection of the 125 GeV scalar boson at the LHC.
In EW baryogenesis, the condition of departure from thermal equilibrium is realized by a strong first-order  phase transition (SFOPT).
We focus our study on the condition of SFOPT, which could also produce detectable gravitational waves at future gravitational wave detectors by bubble wall collisions, magnetohydrodynamic turbulence of bubbles, and sound waves in the hot plasma of the early universe~\cite{Witten:1984rs, Hogan:1984hx, Turner:1990rc,Kamionkowski:1993fg,Kosowsky:2001xp,Caprini:2009yp,Hindmarsh:2013xza,Hindmarsh:2015qta}.
Unfortunately,  a 125 GeV Higgs mass is too heavy to realize a SFOPT~\cite{Morrissey:2012db} in the Standard Model (SM). Thus, the Higgs sector is extended in order to produce a SFOPT in many new physics (NP) models.

The SFOPT has been classified into four categories in Ref.~\cite{Chung:2012vg}.
One interesting class is the type of SFOPT from the tree-level barrier induced by non-renormaliazble operators.
In this work, we focus our attention on this type of SFOPT, induced by the dimension-6 (dim-6) effective operator $|H|^6$~\cite{,Grojean:2004xa,Zhang:1992fs} in the framework of effective field theory (EFT)~\cite{Zhang:1992fs,Zhang:1993vh,Whisnant:1994fh,Zhang:1994fb,Huang:2015bta,Kobakhidze:2015xlz,Balazs:2016yvi}.
The $|H|^6$ operator can be generated by heavy particles or strong dynamics at some high scale (at which the excitations of the underlying theory can be directly probed).
Many other operators also appear at the same time. A full effective Lagrangian at  dim-6 level is
\begin{eqnarray}
\mathcal{L} &\supset& -\mu^2|H|^2-\lambda|H|^4+c_6|H|^6+\sum_i c_i \CO_i,
\end{eqnarray}
where $\CO_i$ denotes other dim-6 operators and the coefficients $c_{i}$'s have dimensions of $[{\rm GeV}]^{-2}$. Note that the coefficients are no longer independent in a given NP theory.
Hence, one must consider the complete set of relevant dim-6 operators when studying the EW phase transition and collider phenomenology.

In this work we adopt the so-called covariant derivative expansion (CDE) method~\cite{Henning:2014wua} to derive these dim-6 operators and their Wilson coefficients in three NP models with scalar extensions, including the triplet extension model, the doublet extension model, and the singlet extension model. We further demonstrate that: i) there is plenty of parameter space satisfying both the SFOPT condition and current experimental data when considering all the dim-6 operators; and ii) one can explore this type of SFOPT scenario through $Zh$ production at future lepton colliders, e.g. the Circular Electron-Positron Collider (CEPC)~\cite{CEPC-SPPCStudyGroup:2015csa}, the International Linear Collider (ILC)~\cite{Gomez-Ceballos:2013zzn}, and the Future Circular Collider (FCC-ee)~\cite{dEnterria:2016fpc, d'Enterria:2132590}.

The paper is organized as follows. In Section~\ref{sec:EFT}, we describe the effective operators in the EFT framework, and show that the dim-6 operators can change the Higgs potential, realize the SFOPT, and make contributions to the EW observables and $Zh$ cross section. In Section~\ref{sec:cons}, we discuss constraints on all the dim-6 operators from SFOPT and the EW precision test in three scalar extension models. Predictions for the $Zh$ cross section deviation are also given. Finally, we conclude in Section~\ref{sec:sum}.

\section{Effective theory for electroweak phase transitions}\label{sec:EFT}

We adopt the EFT approach~\cite{Georgi:1991ch,Weinberg:1978kz,Wudka:1994ny} to study the SFOPT. Let $\Lambda$ be the NP scale at which the excitations of the underlying theory can be probed directly. If $\Lambda$ is so high that none of the heavy excitations can be directly produced, all NP effects can be parameterized by gauge-invariant operators constructed out of the SM fields. Those operators are high dimensional, and suppressed by inverse powers of $\Lambda$. The dim-5 operators violate lepton
number~\cite{Weinberg:1979sa,Wilczek:1979hc,Weldon:1980gi},
and are bounded strongly by existing data~\cite{Buchmuller:1985jz};
the largest contributions are then expected to be generated by dim-6 operators, which are denoted as $\CO_{i}$.
The effective Lagrangian can then  be expressed as
\begin{eqnarray}
\mathcal{L}_{\rm eff}&=&\mathcal{L}_{\rm SM}+\frac{1}{\Lambda^2}
\sum_i\left(\mathcal{C}_i\CO_i+ \hbox{H.c.} \right)+O\left(\Lambda^{-3}\right)\nn\\
&=& \mathcal{L}_{\rm SM}+
\sum_i\left(c_i\CO_i+ \hbox{H.c.} \right)+O\left(\Lambda^{-3}\right),
\end{eqnarray}
where $c_i\equiv \mathcal{C}_i/\Lambda^2$ are coefficients that parametrize the non-SM
interactions and are to be determined by matching the full theory to the effective operators at the scale $\Lambda$.

To obtain the effective operators in a given NP model, we apply the CDE procedure, which computes the path integral of a heavy field $\Psi$ by use of the saddle point method~\cite{Henning:2014wua}. We expand the complete action around the classical solution of $\Psi$ (determined by the configuration of SM fields), with the leading term (i.e. tree level) being the classical action and the next-to-leading term (i.e. one-loop level) being a Gaussian integral. Thus, we obtain a set of effective operators when the heavy field $\Psi$ decouples.

The relevant operators for our study are separated into two categories as follows~\cite{Falkowski:2014tna}:
\begin{itemize}
\item the bosonic operators:
\begin{align}
\label{table:one1}
\mathcal{L}_{\rm eff}^{\rm b}&=\frac{c_H}{2}(\partial^\mu |H|^2)^2+\frac{c_T}{2}\left (H^\dagger \Dfbd H\right)^2+c_6|H|^6\nn\\
&+c_Wig\left( H^\dagger  \sigma^a \Dfbd H \right )D_\nu  W^{a\mu \nu}\nn \\
&+c_Big' \left( H^\dagger  \Dfbd H \right )\partial_\nu B^{\mu \nu}\nn \\
&-\frac{c_{2W}}{2}  ( D^\mu  W_{\mu \nu}^a)^2-\frac{c_{2B}}{2}( \partial^\mu  B_{\mu \nu})^2\nn\\
&+c_{WW}g^2 |H|^2 W^a_{\mu\nu}W^{a \mu\nu}+c_{BB}g^{\prime 2} |H|^2 B_{\mu\nu}B^{\mu\nu}\nn\\
&+c_{WB}g{g}^{\prime} H^\dagger \sigma^a H W^a_{\mu\nu}B^{\mu\nu}\nn\\
&+c_{HW}ig (D^\mu H)^\dagger \sigma^a (D^\nu H) W^a_{\mu\nu}\nn\\
&+c_{HB}ig^\prime(D^\mu H)^\dagger  (D^\nu H) B_{\mu\nu} \nn\\
&+\frac{c_{3W}}{3!} g\epsilon_{abc}W^{a\, \nu}_{\mu}W^{b\,\rho}_{\nu}W^{c\, \mu}_\rho;
\end{align}
\item the fermion-involved operators:
\begin{align}
\mathcal{L}_{\rm eff}^{\rm f}&=c_L^f(iH^\dagger\Dfbd H)(\bar f_L\gamma^\mu f_L)\nn \\
&+c_L^{(3)f}(iH^\dagger\sigma^a\Dfbd H)(\bar f_L\gamma^\mu\sigma^a f_L)\nn \\
&+c_R^f(iH^\dagger\Dfbd H)(\bar f_R\gamma^\mu f_R)\nn\\
&+c_{LL}^{(3)ff'}(\bar f\gamma_\mu\sigma^a f_L)(\bar f'_L\gamma^\mu\sigma^a f'_L)\nn\\
&+c_y^u|H|^2\bar q_L \tilde H u_R+c_y^d|H|^2\bar q_L H d_R+c_y^e|H|^2\bar \ell_L H e_R,
\label{table:one11}
\end{align}
\end{itemize}
where
\begin{align}
&D_{\rho} W^a_{\mu \nu} = \partial_\rho W^a_{\mu \nu} + g \epsilon^{abc} W^b_\rho W^c_{\mu\nu},\nn\\
&D_\mu H = \partial_\mu H -i g\frac{\sigma^a}{2} W^a_\mu H - i g' Y_H B_\mu H,\nn\\
& H^\dagger \Dfbd H\equiv H^\dagger D_\mu H - (D_\mu H)^\dagger H,
\end{align}
and $\sigma^a$ are the Pauli matrices.
The hypercharge of the Higgs boson is chosen as $Y_H=1/2$.

It is known that the dim-6 operators can be transformed into each other by use of the Fierz identity and the equation of motion (EOM). For example, the $\CO_{2W}$ and $\CO_{2B}$ operators can be reduced to $\CO_H$, $\CO_T$, $\CO_L$, $\CO_L^{(3)}$, $\CO_R$ and four-fermion operators by using the EOM of gauge field,
\be\begin{split}
D^\nu W^a_{\mu\nu}&=igH^\dagger\frac{\sigma^a}{2}\Dfbd H+g\sum_f\bar f_L\frac{\sigma^a}{2}\gamma_\mu f_L,\nn\\
\partial^\nu B_{\mu\nu}&=ig'Y_HH^\dagger\Dfbd H+g'\sum_f(Y_L^f\bar f_L\gamma_\mu f_L+Y_R^f\bar f_R\gamma_\mu f_R),\nn
\end{split}\ee
where $Y^f_L$ and $Y^f_R$ denote the hypercharges of left-handed and right-handed fermion fields, respectively. In addition, making use of the EOM we can get the identities~\cite{Elias-Miro:2013mua}
\begin{align}
&\CO_W=\mathcal{O}_{HW}+\frac{1}{4}(\mathcal{O}_{WW}+\mathcal{O}_{WB}),\nn\\
&\CO_B=\mathcal{O}_{HB}+\frac{1}{4}(\mathcal{O}_{BB}+\mathcal{O}_{WB}),
\end{align}
and
\begin{align}
\CO_W=&g^2\Big[-\frac{3}{2}\CO_H+2\lambda\CO_6+\frac{1}{2}(y_u\CO_y^u+y_d\CO_y^d+y_e\CO_y^e+\text{h.c.})\nn\\
&+\frac{1}{4}\sum_f\CO_L^{(3)f}\Big],\nn\\
\CO_B=&g'^2\Big[-\frac{1}{2}\CO_T+\frac{1}{2}\sum_f\big(Y_L^f\CO_L^f+Y_R^f\CO_R^f\big)\Big],
\end{align}
where $\lambda$ denotes the quartic coupling of the Higgs field, while $y_f$ denotes the Yukawa coupling of the fermion fields. Clearly, such equations imply that Eqs.~(\ref{table:one1}) and~(\ref{table:one11}) form a redundant set of operators.

Those independent operators are defined as the ``basis'' in the EFT description.
There are 59 independent baryon number conserving dim-6 operators. Under the simplification of just one fermion generation, 76 independent real parameters are needed to describe the effects of the above operators. However, when all three generation fermions and the flavor structures are taken into account, the number of independent parameters increases remarkably: there will be in total 2499 independent parameters (see Refs.~\cite{Alonso:2013hga, Brivio:2017btx}). Several bases have been proposed to characterize different types of NP or to serve for different phenomenological studies. One is called the ``Warsaw basis''~\cite{Grzadkowski:2010es}, and is obtained by eliminating $\CO_W$, $\CO_B$, $\CO_{2W}$, $\CO_{2B}$, $\CO_{HW}$ and $\CO_{HB}$. For EW precision and Higgs phenomenology, there are several other convenient bases, e.g. the HISZ (Hagiwara) basis~\cite{Hagiwara:1993ck}, the SILH (strongly-interacting light Higgs) basis~\cite{Giudice:2007fh,Elias-Miro:2013mua,Pomarol:2013zra}, and the EGGM basis~\cite{Elias-Miro:2013eta}. They mainly differ in the choice of bosonic operators. For example, the SILH basis can be achieved by dropping $\CO_{WW}$, $\CO_{WB}$, $\CO_L^\ell$, $\CO_L^{(3)\ell}$ and some four-fermion operators.

Finally, we comment on the operator $\CO_r$~\cite{Henning:2014wua},
\be
\CO_r=|H|^2|D_\mu H|^2,
\ee
which can be transformed into the operators in Eqs.~(\ref{table:one1}) and~(\ref{table:one11}) by the EOM of the Higgs boson field:
\begin{align}
\label{O_r}
\CO_r =&\mu^2|H|^4-\CO_H+2\lambda\CO_6+\sum_u\frac{y_u}{2}(\CO_y^u+\text{h.c.})\nn\\
&+\sum_d\frac{y_d}{2}(\CO_y^d+\text{h.c.})+\sum_e\frac{y_e}{2}(\CO_y^e+\text{h.c.}).
\end{align}
Here, $\mu^2$ and $\lambda$ denote the quadratic and quartic coupling of the Higgs field, respectively.

\subsection{Strong first-order phase transition}

The dim-6 operator $|H|^6$ in the Higgs potential,
\begin{equation}\label{v00}
V(H)=\mu^2 |H|^2 + \lambda |H|^4-c_6 |H|^6,
\end{equation}
introduces a tree-level barrier so as to realize the SFOPT~\cite{Zhang:1992fs,Zhang:1993vh,Whisnant:1994fh,Zhang:1994fb,Huang:2015bta}; see Refs.~\cite{Grojean:2004xa, Delaunay:2007wb, Grinstein:2008qi,Chung:2012vg,Ham:2004zs,Bodeker:2004ws,Chu:2015nha,Huang:2016odd} for recent studies.
Note that the values of $\mu^2$ and $\lambda$ are no longer the SM values, in order to satisfy the SFOPT condition  and give a 125.09 GeV Higgs boson.
The contributions from other dim-6 operators to the EW phase transition are negligible, for the following reasons.
%The contributions of the dim-6 operators are greatly suppressed by the
In this type of tree-level barrier SFOPT induced by the potential in Eq.~(\ref{v00}), the $c_6 |H|^6$ term directly contributes to
the SFOPT.
Other dim-6 operators contribute to the EW phase transition mainly through modifying the masses of the particles which can contribute to the SFOPT.
However, the mass modifications are negligible from current data,
and the coefficient $c_6$ can be rather large, since
there is nearly no constraint on $c_6$ from current data.
%While the other dim-6 effective operators contribute to the EW phase transition through modifying the masses and couplings of SM particles, which are strongly constrained by the current data.
%Thus, compared to the large $c_6$, the coefficients of the other
%dim-6 operators are small.}

When the SFOPT is considered, one can simplify the potential by substituting $H$ with $h/\sqrt{2}$:
\begin{equation}\label{v0}
V_{\rm tree}(h) = \frac{1}{2}\mu^2 h^2 + \frac{\lambda}{4} h^4 -\frac{c_6}{8} h^6.
\end{equation}
The corresponding finite-temperature effective potential up to one-loop level can be written as~\cite{Quiros:1999jp,Dolan:1973qd},
\begin{equation}\label{fullpotential}
 V_\mathrm{eff}(h,T)=V_\text{tree}(h)+
V_1^{T=0}(h)+\Delta V_1^{T\neq 0}(h,T),
\end{equation}
where $V_\text{tree}(h)$ is the tree-level potential in Eq.~(\ref{v0}), $V_1^{T=0}(h)$ is the Coleman-Weinberg
potential at zero temperature, and $\Delta V_1^{T\neq 0}(h)$ represents the leading thermal effects with daisy resummation.
After including the full one-loop results given in Refs.~\cite{Delaunay:2007wb,Bodeker:2004ws},
the washout condition for the SFOPT, $v(T_c)/T_c \gtrsim 1$,
can  easily be satisfied~\cite{Bodeker:2004ws}.
More precise washout conditions, based on a detailed study on the sphaleron process
with the dim-6 effective operators, are given in
Refs.~\cite{Spannowsky:2016ile,Gan:2017mcv}.
If the EW phase transition is a SFOPT, then, at one-loop level,
the SFOPT and vacuum stability give the following constraints~\cite{Huang:2015izx}
\begin{equation}\label{c6}
\frac{1}{(0.89 \rm~TeV)^2} <- c_6 <\frac{1}{(0.55~\rm TeV)^2}.
\end{equation}
The SFOPT condition modifies the trilinear Higgs boson as follows:
\begin{equation}\label{3Hvtx}
\mathcal{L}_{hhh}= -\frac{1}{3!} (1+ \delta_h) A_{h} h^3,
\end{equation}
where $A_{h}=3 m_h^2/v$ is the trilinear Higgs boson coupling in the SM,
and $\delta_h$ is the modification of the trilinear Higgs coupling induced by the dim-6 operator. In this scenario,
\be
\delta_h \approx -0.468c_6 \times\text{TeV}^2,
\ee
and roughly varies from $0.6$ to $1.5$ in the allowed parameter space.
The $|H|^6$ operator yields a distinctive signal at the Large Hadron Collider (LHC), e.g. two peaks in the invariant mass distribution of the Higgs boson pairs~\cite{Huang:2015bta}.
Unfortunately, due to its low experimental precision, the LHC is not capable of testing this type of EW phase transition scenario.
However, precise information on the triplet Higgs coupling might be obtained at future lepton colliders~\cite{Noble:2007kk,Katz:2014bha,Curtin:2014jma}~\footnote{Gravitational waves experiments can provide a complementary approach to testing the EW phase transition~\cite{Huang:2015izx, Huang:2016odd,Cai:2017tmh}.}.

\subsection{Electroweak precision tests}

In order to describe EW observables, we use the $Z$-scheme in which three of the most precisely measured values \{$\alpha$, $m_Z$, $G_F$\} are chosen as the input parameters. Other EW observables,  e.g. $m_W$ and $\Gamma_{Z}$, are expressed as a function of the three input parameters (together with Higgs mass or fermion masses, if necessary).
We first consider the EW precision measurements at the Tevatron and the Large Electron-Positron Collider (LEP), including
\be\label{EW_observables}
m_W, ~N_\nu, ~A_{FB}^b, ~R_b, ~R_\mu, ~R_\tau, ~\sin^2\theta^l_{\rm eff}.
\ee
We check that, for the NP models considered here, the constraints from triple gauge coupling measurements at the LEP and LHC are quite weak. Thus, we do not consider these here.
The scalar extended models can have a remarkable impact on Higgs-relevant processes since they often have sizable interactions with the Higgs field. Therefore, we also consider the measurement of  Higgs decay branching ratios at the LHC, e.g.
\begin{eqnarray}
\label{Higgs_data}
&&\text{Br}_{h\to WW}, \quad \text{Br}_{h\to ZZ}, \quad \text{Br}_{h\to\gamma\gamma},\nn\\
&&\text{Br}_{h\to gg}, \qquad \text{Br}_{h\to\tau\tau},~\quad  \text{Br}_{h\to\mu\mu}.
\end{eqnarray}
In comparison with the low energy precision observables listed in Eq.~(\ref{EW_observables}), the Higgs data impose rather weak constraints on NP model parameters, but they can offer constraints on some specific operators that are nearly free of EW precision measurements. For example, at tree level, the contributions of $\CO_{WW}$ and $\CO_{BB}$ to EW observables can be absorbed into the redefinition of gauge boson field and gauge coupling, leaving no physics effects. Such operators can be probed only through Higgs-involved process. Their contributions to EW observables are through Higgs loops, and hence are tiny due to loop suppression. However, the observables listed in Eq.~(\ref{Higgs_data}) can probe such operators at tree level.

\subsection{Phenomenology at future lepton colliders}

One way to test the EW phase transition scenario at future lepton colliders is through the $Z$-boson and Higgs boson associated production (the $Zh$ channel)~\cite{McCullough:2013rea,Englert:2013tya}.
The cross section of the $Zh$ channel $\sigma(Zh)$ could be measured with an accuracy of $\mathcal{O}(0.1 \%)\sim \mathcal{O}(1\%)$ at future lepton colliders \cite{Baer:2013cma,Azzi:2014jwa}.
For example, an accuracy of $0.5\%$ for $\sigma_{Zh}$ measurement could be achieved at the CEPC with an integrated luminosity of $5~\mathrm{ab}^{-1}$~\cite{CEPC-SPPCStudyGroup:2015csa,Ruan:2014xxa}, while the FCC-ee is expected to do a better job~\cite{dEnterria:2016fpc, d'Enterria:2132590}.
The operator $\CO_6=|H|^6$ contributes to the $Zh$ cross section through a triangle loop.
Other dim-6 operators, e.g. $\CO_H = \frac{1}{2} (\partial_\mu |H|^2)^2$ and $\CO_T = \frac{1}{2}(H^\dagger \Dfbd H)^2$, contribute to the $Zh$ production at tree-level.
We define the deviation of cross section of the $Zh$ production, normalized to the SM cross section, as follows:
\begin{equation}
\delta_{\sigma(Zh)}\equiv \frac{\sigma_{Zh}}{\sigma_{Zh}^\mathrm{SM}}-1.
\end{equation}
As the golden channel of Higgs physics, the $Zh$ production rate has been calculated very precisely, including one-loop and two-loop quantum corrections~\cite{Huang:2015izx, Gong:2016jys,Sun:2016bel}. Here, we refer $\sigma_{Zh}^{\rm SM}$ to the most state-of-art calculation of $Zh$ production in the SM.

At a lepton collider with $\sqrt{s}=250~{\rm GeV}$ ($\sqrt{s}$ being the center of mass energy of the collider), the high dimension operators' contribution to the $Zh$ production is approximately given by~\cite{McCullough:2013rea,Englert:2013tya}
\begin{align}
\label{zh_deviation}
\delta_{\sigma(Zh)} \simeq & ( 0.26 c_{WW} + 0.01 c_{BB} +0.04 c_{WB} \nn\\&-0.06 c_H
 -0.04 c_T +0.74 c_L^{(3)\ell}  + 0.28 c_{LL}^{(3)\ell}\nn\\ & + 1.03 c_L^\ell -0.76 c^e_R) \times\text{TeV}^2+0.016 \delta_h,
\end{align}
where $\delta_h$ is the deviation of the Higgs trilinear coupling defined in Eq.~(\ref{3Hvtx}).
The $\delta_h$ contribution, suffering from a loop suppression, is usually ignored in the operator analysis of  the $Zh$ channel~\cite{Craig:2014una, Ge:2016zro}.
However, we argue that it cannot be ignored in our study of the SFOPT, owing to the following reasons. First, the SFOPT condition requires a large $c_6$ coefficient, resulting in a considerable contribution to $\delta_{\sigma(Zh)}$ of $0.96\%\sim2.4\%$. Note that there are nearly no constraints on $c_6$ from current experiments. Second, the constraints on the coefficients of other dim-6 operators are stronger than $c_6$.
Hence, the $\delta_h$ contribution cannot be ignored.

\section{New physics models and dim-6 operators}\label{sec:cons}

In this section we consider three NP models with scalar extensions and use the CDE method to derive the full set of dim-6 operators in the three NP models. The impact of those operators on the EW phase transition is examined. After considering the LEP and LHC constraints we explore the potential of observing the effects of those operators in the $Zh$ channel at future lepton colliders.

\subsection{Model with $SU(2)_L^{}$ triplet scalar}

We first consider a weak $SU(2)_L$ triplet scalar extension model. For simplicity, we choose the triplet scalar not gauging under the $U(1)_Y$ group. We define the triplet scalar $\Sigma$ as
\begin{align}
\begin{array}{l}\Sigma(1,3,0)\end{array}\!\!&=&\left[\begin{array}{cc}\frac{1}{2}\delta^{0}_{}
&\frac{1}{\sqrt{2}}\delta^{+}_{}\\
[2mm]
\frac{1}{\sqrt{2}}\delta^{-}_{}&-\frac{1}{2}\delta^{0}_{}\end{array}\right]=\Sigma^\dagger_{}=\Sigma^a \frac{\sigma^a}{2}.
\end{align}
The relevant Lagrangian involving the triplet scalar $\Sigma$ is
\begin{eqnarray}
\label{tl1}
\delta\mathcal{L}&=&\textrm{Tr}[(D^\mu_{}\Sigma)^\dagger_{}D_\mu^{}\Sigma]-M_{\Sigma}^2\textrm{Tr}(\Sigma^2_{})-\zeta_\Sigma^{}[\textrm{Tr}(\Sigma^2_{})]^2_{}\nn\\
&&+2\xi^{}_\Sigma H^{\dagger}\Sigma H-2\kappa_\Sigma^{}|H|^2\textrm{Tr}(\Sigma^{2}_{})\,,
\end{eqnarray}
where the covariant derivative is given by $D_\mu \Sigma=[D_{\mu},\Sigma]$.
The triplet scalar exists in many NP models, e.g.  Grand Unification Theories, Little Higgs models,  neutrino mass models, the Georgi-Machacek model~\cite{Georgi:1985nv}, and the $SU(3)_C\otimes SU(3)_L\otimes U(1)_X$ gauge extension model~\cite{Cao:2015scs,Cao:2016uur,Huang:2017laj}.
We keep $\zeta_\Sigma>0$ and $\kappa_\Sigma>0$, so that the scalar potential is bounded below.

\begin{figure*}
  \centering
    \includegraphics[scale=0.3]{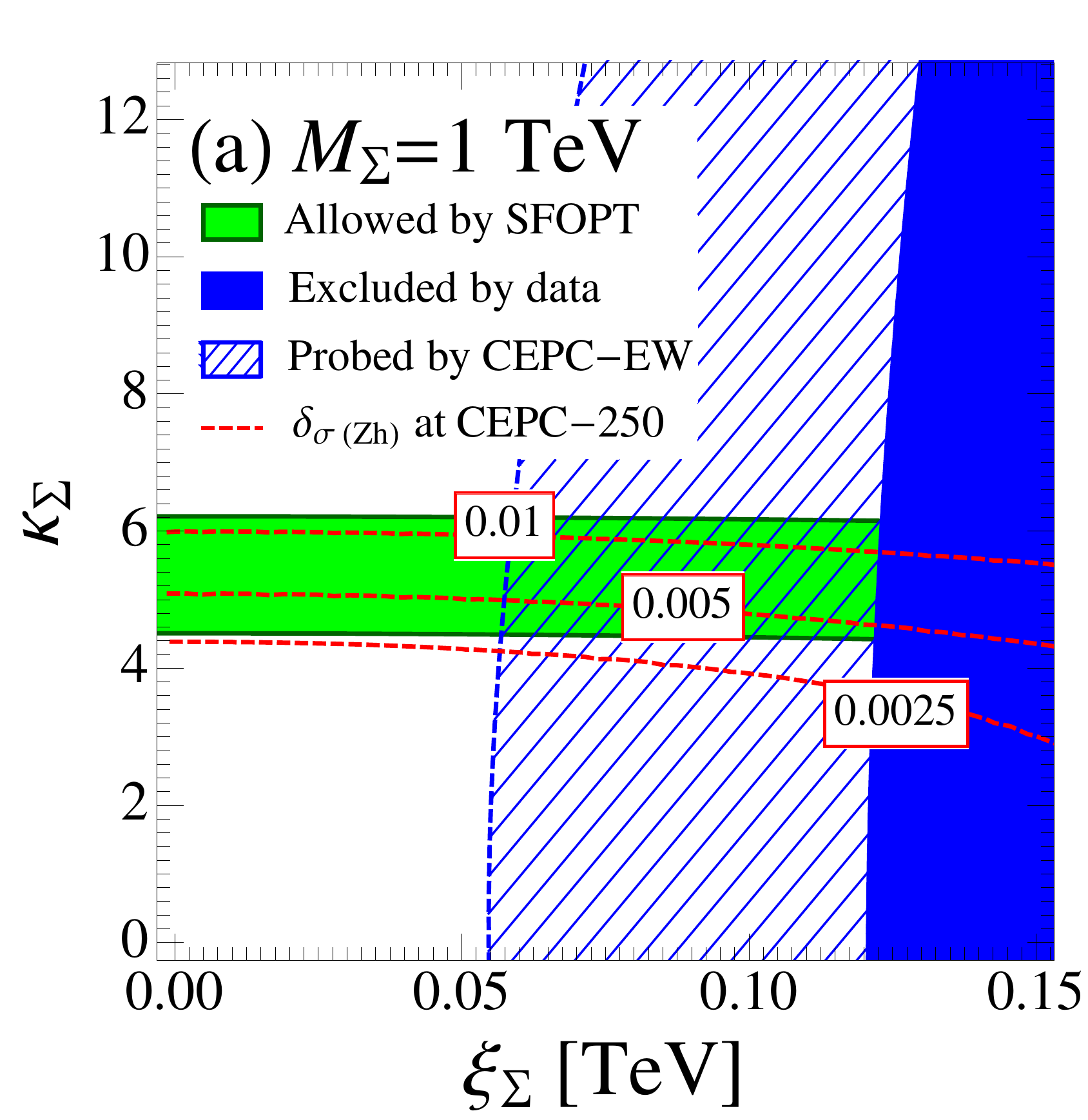}
    \includegraphics[scale=0.3]{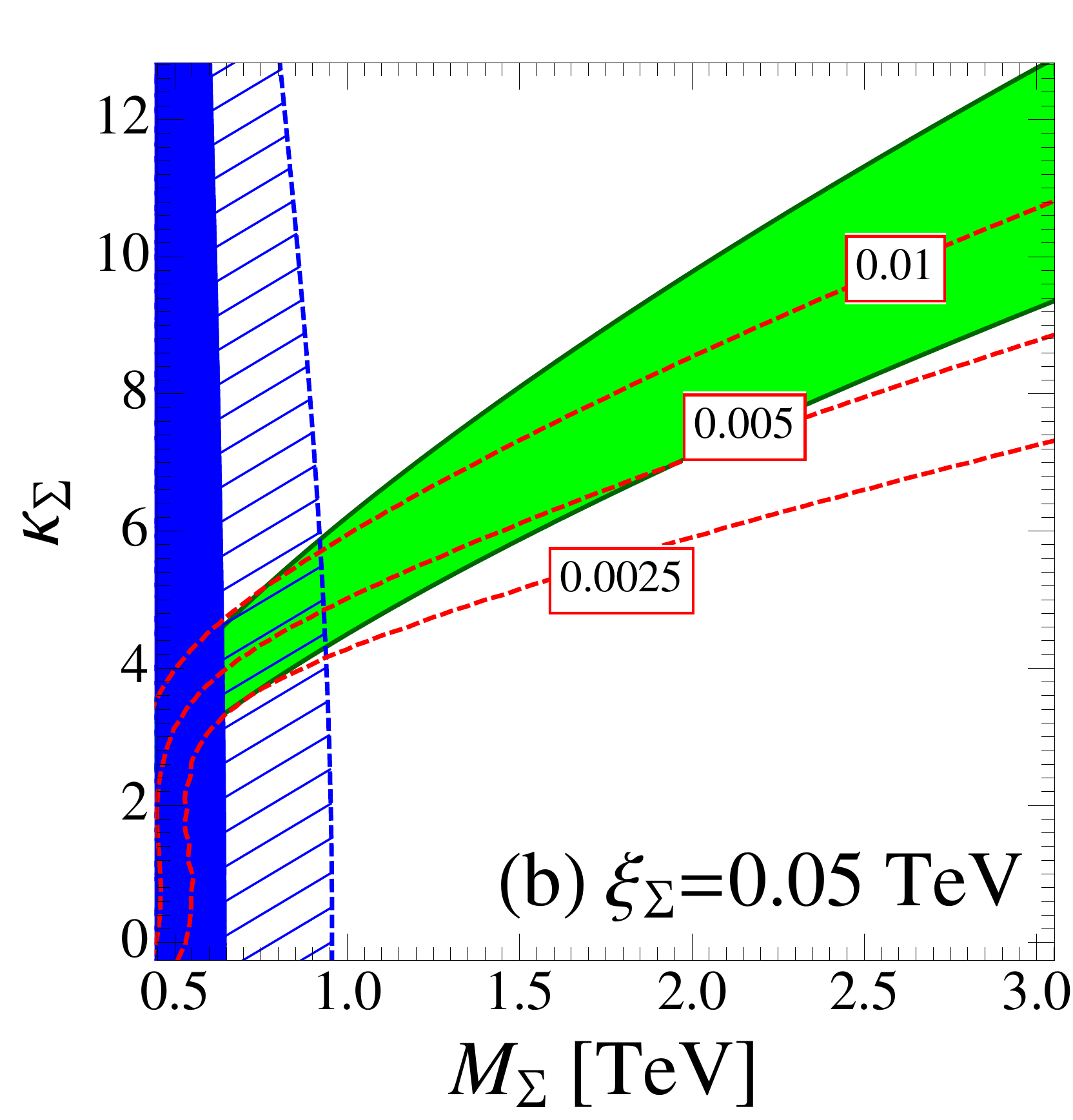}
    \includegraphics[scale=0.3]{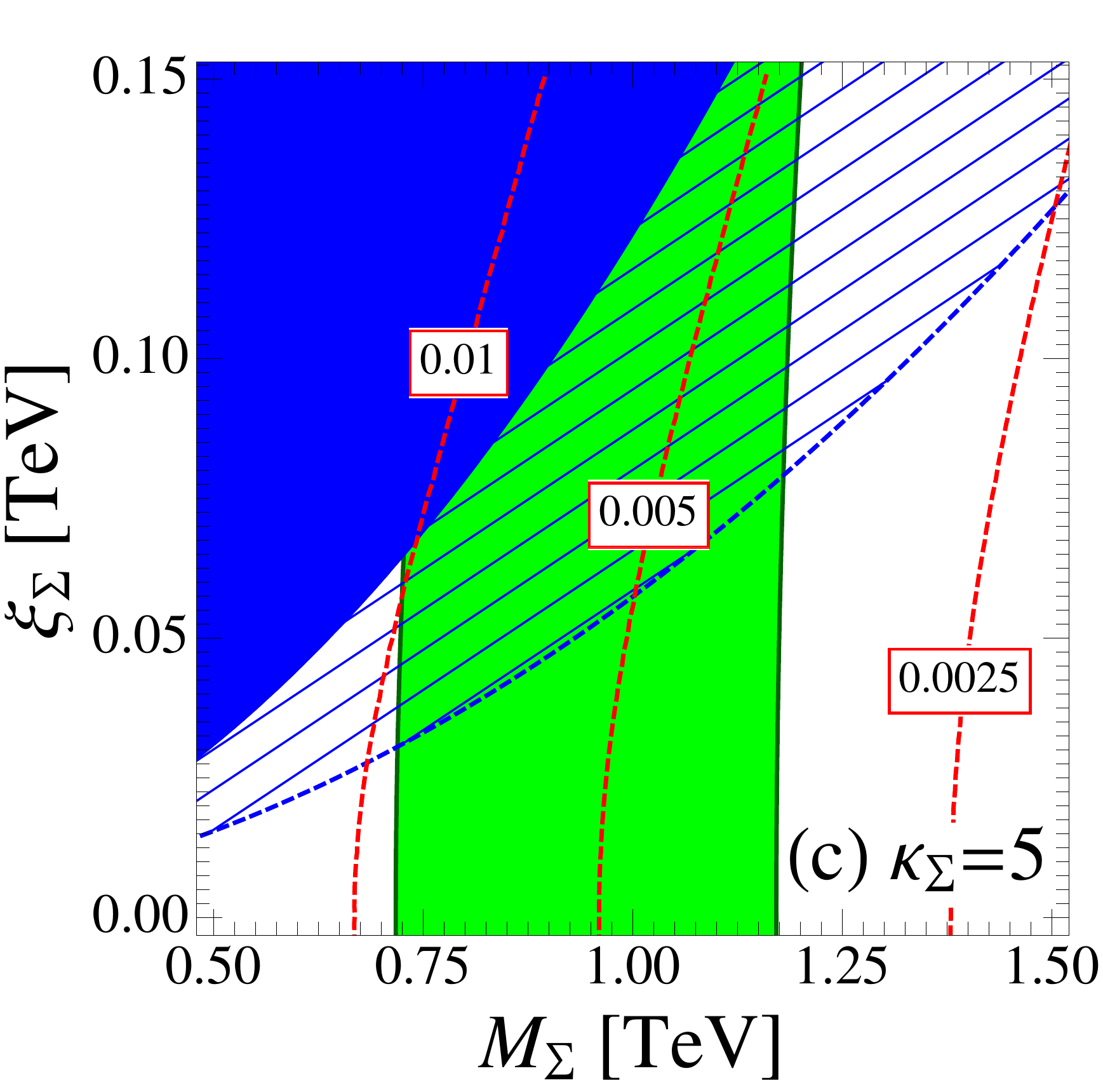}
\caption{The parameter space of a triplet model compatible with SFOPT (green) and current EW and Higgs boson data (blue). The hatched region can be covered by the CEPC, while the red curves denote $\delta_{\sigma(Zh)}$ at the CEPC.}
\label{triplet_no_hypercharge}
\end{figure*}

Assuming the triplet scalar decouples, we use the CDE method to integrate out the $\Sigma$ field. We obtain the coefficients of those dim-6 operators mentioned in Section~\ref{sec:EFT} as follows:
\begin{align}
c_{WW}&=\frac{1}{(4\pi)^2 }\frac{\kappa_\Sigma}{6M_\Sigma^2}, \nn\\
c_{2W}&=c_{3W}=\frac{1}{(4\pi)^2 }\frac{g^2}{30 M_\Sigma^2},\nn\\
c_{H}&=\frac{1}{(4\pi)^2 }\frac{\kappa_\Sigma^2}{M_\Sigma^2},\nn\\
c_{T}&=\frac{ \xi_\Sigma^2}{M_\Sigma^4}+\frac{1}{(4\pi)^2}\frac{10\zeta_\Sigma\xi_\Sigma^2}{M_\Sigma^4}, \nn\\
c_r&=\frac{2\xi_\Sigma^2}{M_\Sigma^4}+\frac{1}{(4\pi)^2}\frac{20\zeta_\Sigma\xi_\Sigma^2}{M_\Sigma^4},\nn\\
-c_6&=\frac{\kappa_\Sigma \xi_\Sigma^2}{M_\Sigma^4}+\frac{1}{(4\pi)^2 }\frac{2\kappa_\Sigma^3}{M_\Sigma^2}+\frac{1}{(4\pi)^2}\frac{10\zeta_\Sigma\kappa_\Sigma\xi_\Sigma^2}{M_\Sigma^4},
\label{tab:hit}
\end{align}
where those terms without (with) the factor of $1/(4\pi)^2$ are induced by the tree (one-loop) level contributions, respectively. The coefficients are highly correlated to respect the weak quantum number of the triplet $\Sigma$.  Our results are consistent with Refs.~\cite{Henning:2014wua, Khandker:2012zu}.  The operator $\CO_r$ can be reduced to $\CO_H$ and $\CO_6$ by using Eq.~(\ref{O_r}). Note that the Wilson coefficients are determined completely by the model parameters $\kappa_\Sigma$, $\xi_\Sigma$, $\zeta_\Sigma$ and $M_\Sigma$. We require
\be
\kappa_\Sigma<4\pi,\quad \zeta_\Sigma<4\pi,\quad \frac{\xi_\Sigma}{M_\Sigma}<4\pi,
\ee
to ensure perturbativity and unitarity~\cite{Khan:2016sxm}.

For simplicity, we set $\zeta_{\Sigma}=0$. Given the tiny coefficients of $\CO_{2W}$ and $\CO_{3W}$, their effects are negligible and we can omit them safely. Because the operators are induced at the scale of $M_\Sigma$, when we use them to discuss EW scale physics, the effects of renormalization group (RG) running should be studied carefully. After converting the operators into the Warsaw basis,
following Refs.~\cite{Jenkins:2013zja, Jenkins:2013wua, Alonso:2013hga} we calculate the RG running effects and obtain the dim-6 operators and their Wilson coefficients at the EW scale.
The condition of SFOPT is identified as
\begin{equation}
\frac{1}{(0.89 \rm~TeV)^2} < -c_6(\kappa_\Sigma,\xi_\Sigma,M_\Sigma) \big|_{\text{EW scale}}<\frac{1}{(0.55~\rm TeV)^2}.
\end{equation}

Armed with the Wilson coefficients of dim-6 operators at the weak scale, we perform a global $\chi^2$ fit to obtain constraints on the dim-6 operators from the EW precision test and Higgs branching ratio measurements; see Eqs.~(\ref{EW_observables}) and (\ref{Higgs_data}).
The dependence of EW precision observables on the dim-6 operators is calculated in Ref.~\cite{Falkowski:2014tna, Elias-Miro:2013mua, Pomarol:2013zra}. The SM predictions of those EW observables are given in Refs.~\cite{Baak:2014ora, Falkowski:2014tna, Heinemeyer:2013tqa}, while the experimental measurements are presented in Refs.~\cite{Olive:2016xmw, ALEPH:2005ab, Group:2012gb,  Khachatryan:2016vau}.
We define $\chi^2$ as
\be
\chi^2=\sum_j\frac{(\hat O_j^{\rm exp}-\hat O^{\rm theo}_j)^2}{(\delta^{\rm exp}_j)^2},
\ee
where $\hat O_j^{\rm exp}$ and $\hat O^{\rm theo}_j$ denote the experimental central value and theoretical prediction of the observable $j$, respectively, while $\delta^{\rm exp}_j$ represents the experimental error of the measurement of observable $j$. $\hat O^{\rm theo}_j$ is a function of the Wilson coefficients $c_i$. As the $c_i$ depend on the model parameters, we derive the exclusion limit on $(\kappa_\Sigma,\xi_\Sigma,M_\Sigma)$ at the 95\% confidence level.
Figure~\ref{triplet_no_hypercharge} shows the allowed parameter space in the plane of ($\kappa_\Sigma$, $\xi_\Sigma$) with $M_\Sigma=1~{\rm TeV}$ (a), ($\kappa_\Sigma$, $M_\Sigma$) with $\xi_\Sigma=0.05~{\rm TeV}$ (b), and ($\xi_\Sigma$, $M_\Sigma$) with $\kappa_\Sigma=5$ (c). The green regions satisfy the SFOPT condition while the blue regions are excluded by the EW  precision test and Higgs branching ratios.
In the model parameter space of interest to us, the heavy scale $M_{\Sigma}\sim (1-3)~{\rm TeV}$.  Running from $M_{\Sigma}$ down to the EW scale ($\sim 100~{\rm GeV}$) modifies the  Wilson coefficients slightly,
therefore, the shape of the green and blue regions can easily be  understood from the Wilson coefficients obtained at the scale of $M_\Sigma$; see Eq.~(\ref{tab:hit}).

The constraint from the EW precision test is  predominantly from the oblique $T$-parameter (or equivalently, the $\rho$-parameter). The $\rho$-parameter characterizes the weak isospin breaking that cannot be accounted for by the SM Higgs doublet, and has been measured very precisely, e.g. $\rho=1.00037\pm 0.00023$~\cite{Olive:2016xmw}. As the operator $\CO_T$ explicitly breaks the weak isospin, its coefficient $c_T\sim\xi_\Sigma^2/M_\Sigma^4$ is severely constrained, yielding a tiny value of $\xi_\Sigma$. For example, $\xi_\Sigma\leq 0.1~{\rm TeV}$ for $M_\Sigma=1~{\rm TeV}$. Moreover, the $c_T$ has nothing to do with $\kappa_\Sigma$, therefore, the EW precision test is not sensitive to $\kappa_\Sigma$. See the blue bands in Figs.~\ref{triplet_no_hypercharge}(a) and \ref{triplet_no_hypercharge}(b).

The SFOPT condition is controlled solely by $c_6$,
\be
-c_6 \approx \frac{\kappa_\Sigma \xi_\Sigma^2}{M_\Sigma^4}+\frac{1}{(4\pi)^2 }\frac{2\kappa_\Sigma^3}{M_\Sigma^2}.
\ee
Since $\xi_\Sigma$ is tiny in comparison with $M_\Sigma$, the second term $c_6$ dominates over the first term such that the SFOPT condition is not sensitive to $\xi_\Sigma$. See the green bands in Figs.~\ref{triplet_no_hypercharge}(a) and \ref{triplet_no_hypercharge}(c). In order to overcome the $(4\pi)^2$ suppression factor, a large value of $\kappa_\Sigma$ is needed.
Such a large coupling in the triplet extension model may generate a fairly low Landau pole (LP) for the scalar coupling. For example, for the benchmark point
\be\label{benchmark}
\kappa_\Sigma=5,\quad \xi_\Sigma=0.05\text{ TeV,}\quad M_\Sigma=1\text{ TeV},
\ee
the scalar coupling will break the perturbative limit at a scale around $1.1\times10^2$ TeV and blows up at a scale around $2.7\times10^3$ TeV. However, as pointed out in Ref.~\cite{DiLuzio:2015oha}, the SM gauge coupling constants are free of the LP problem for triplet extension models. For a given $\xi_\Sigma$, the SFOPT condition requires $\kappa_\Sigma \propto M_\Sigma^{2/3}$; see Figs.~\ref{triplet_no_hypercharge}(b).

Before upgrading to $\sqrt{s}=250~{\rm GeV}$, the CEPC plans to operate at the EW precision energy (e.g. $Z$-pole and $W$ pair threshold) in its early stages. The EW observables would be measured with an accuracy better than the LEP results. We adopt the CEPC projected uncertainties ($\delta^{\rm exp}_j$) to estimate the sensitivities of the CEPC to the model parameters. If no anomalies are observed in CEPC EW precise operations,  then the hatched regions in Fig.~\ref{triplet_no_hypercharge} are excluded at 95\% confidence level.

The $Zh$ channel can be measured very precisely at the CEPC with $\sqrt{s}=250~{\rm GeV}$; a fraction deviation of cross section larger than $0.51\%$, i.e. $\delta_{\sigma(Zh)}\geq 0.005$, can be detected~\cite{CEPC-SPPCStudyGroup:2015csa}. The dependence of $\delta_{\sigma(Zh)}$ on operators are given in Eq.~(\ref{zh_deviation}).
Those fermionic operators are generated through RG running and their coefficients are very small, e.g.
$c_L^{(3)\ell}\simeq 8\times10^{-5}$ TeV$^{-2}$, $c_L^{\ell}\simeq 2\times10^{-5}$ TeV$^{-2}$ and $c_R^{e}\simeq 2\times10^{-5}$ TeV$^{-2}$ for the choice of Eq.~(\ref{benchmark}).
Numerically, $\delta_{\sigma(Zh)}$ is dominated by $\CO_T$, $\CO_{WW}$ and $\CO_6$.
We plot the contours of $\delta_{\sigma(Zh)}=1\%,~0.5\%,~0.25\%$ in Fig.~\ref{triplet_no_hypercharge}; see the red dashed curves.

Note that the value of $\delta_h$ in the parameter space of SFOPT is about $0.6 \sim 1.5$.  Even though suffering from small coefficient suppression, $\delta_h$ ($\propto c_6$) gives rise to a sizable contribution in $\delta_{\sigma(Zh)}$, similar to other operators. For example,
\be
\delta_{\sigma(Zh)} = \underbrace{0.001}_{\CO_{WW}} -\underbrace{0.009}_{\CO_H}+ \underbrace{0.013}_{\CO_6}= 0.005,
\ee
for the parameter choice of Eq.~(\ref{benchmark}).
Hence, the contour lines of $\delta_{\sigma(Zh)}$ exhibit similar shapes to the SFOPT band. Especially, the red contour lines nearly coincide with the green band in the small $\xi_\Sigma$ region, where $\CO_6$ dominates over  $\CO_T$ such that both $\delta_{\sigma(Zh)}$ and SFOPT are determined by $c_6$.

Figure~\ref{triplet_no_hypercharge} shows that most of the parameter space allowed by the SFOPT condition and EW precision test can be well explored at the 250 GeV CEPC, given that $\delta_{\sigma(Zh)}$ can be measured within an accuracy of 0.51\%~\cite{Ge:2016zro, Ruan:2014xxa}. The precision knowledge of EW observables achieved in the EW precise operation of CEPC can be used to cross-check the $\delta_{\sigma(Zh)}$ measurement, and in addition, it can also probe model parameters insensitive to $Zh$ production.

\subsection{Model with $SU(2)_L$ doublet scalar}

\begin{figure*}
  \centering
    \includegraphics[scale=0.3]{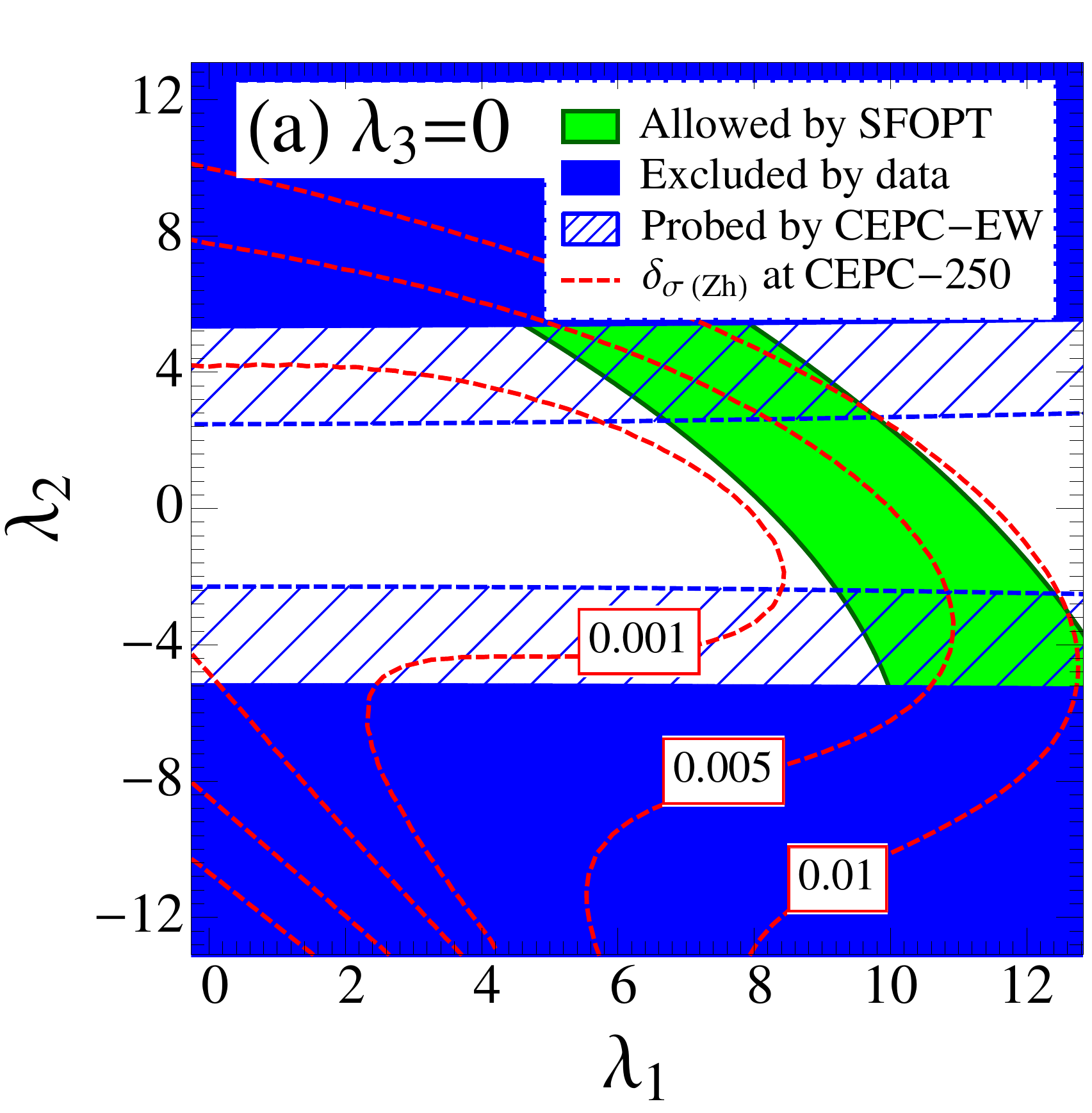}
    \includegraphics[scale=0.3]{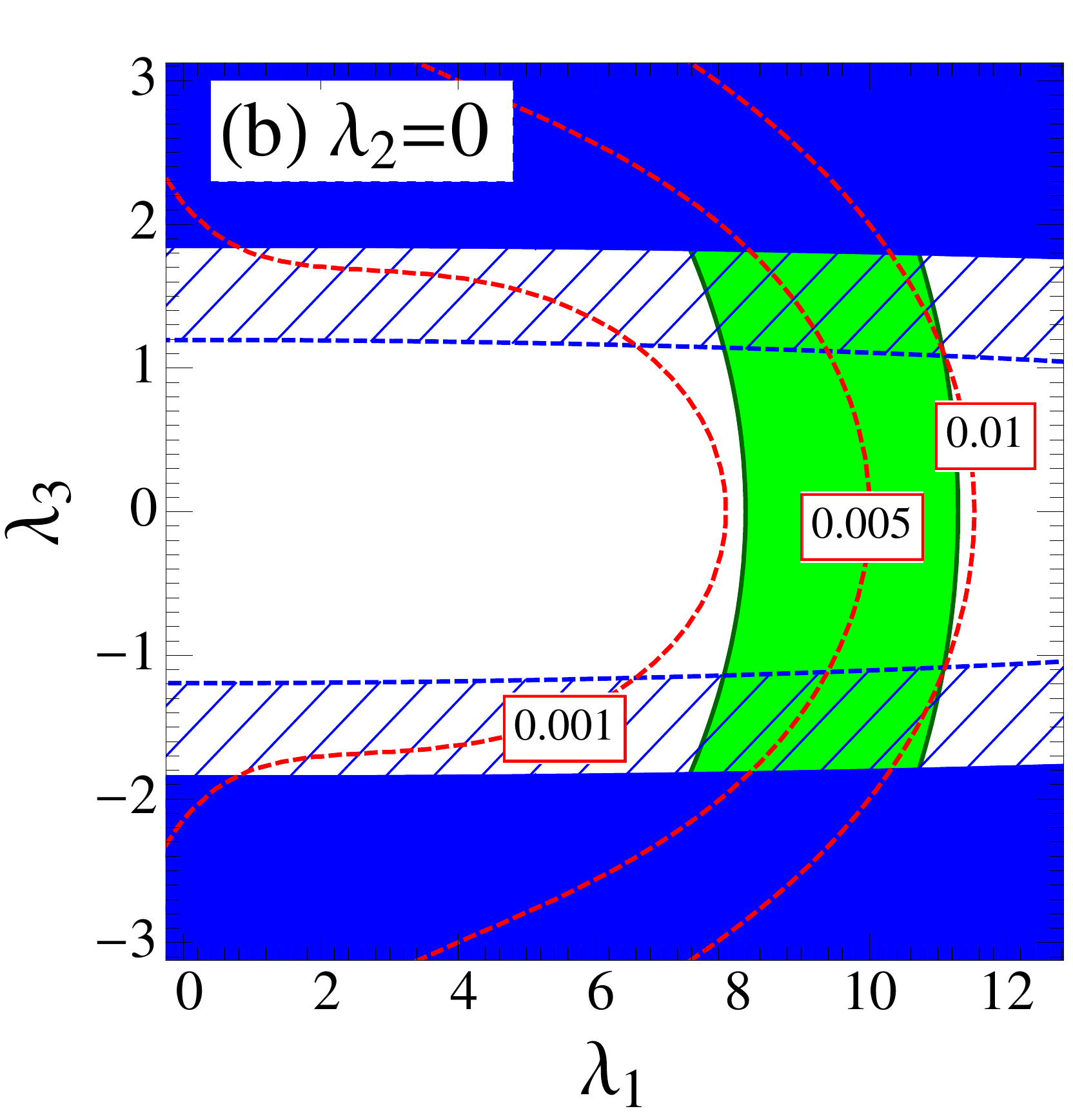}
    \includegraphics[scale=0.3]{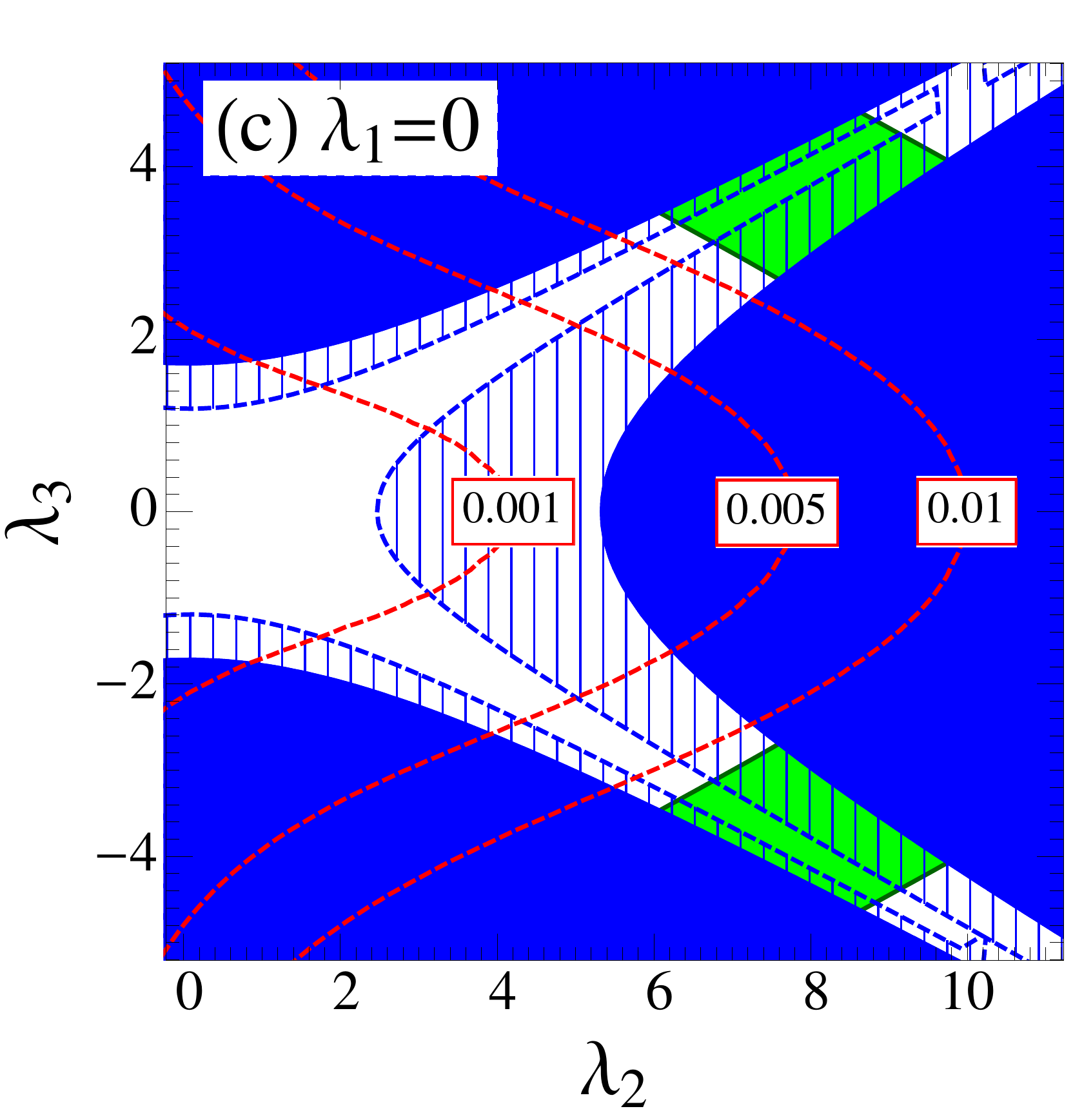}
\caption{
The parameter space of a doublet model compatible with SFOPT (green) and current EW and Higgs boson data (blue) with $M_\Phi=1~{\rm TeV}$. The hatched region can be covered by the CEPC while the red curves denote $\delta_{\sigma(Zh)}$ at the CEPC.
}
\label{doublet_M}
\end{figure*}

The second simplified model we investigated is the doublet extension model. For illustration, we consider a new $SU(2)_L$ doublet scalar $\Phi$ with hypercharge $Y_\Phi=-1/2$. The relevant Lagrangian is
\begin{eqnarray}
\delta \mathcal{L}&=& D_\mu\Phi^\dagger D^\mu \Phi-M_\Phi^2 \Phi^\dag \Phi-\frac{\lambda_{\Phi}}{4}(\Phi^\dag \Phi)^2\nn\\
&&-\lambda_1 \Phi^\dag \Phi H^\dag H-\lambda_2|\Phi \cdot H|^2 -\lambda_3[(\Phi \cdot H)^2+h.c.]\nn\\
&&+(\eta_H |H|^2+\eta_{\Phi} |\Phi|^2)(\Phi\cdot H+h.c.)\,,
\end{eqnarray}
where $\Phi\cdot H=H^T i\sigma^2\Phi$.
The scalar sector shown above mimics the well-studied Two Higgs Doublet Model.
When $M_\Phi\gg v$, $\Phi$ can be integrated out to induce dim-6 operators. The Wilson coefficients of dim-6 operators matched at the scale $M_\Phi$ in the doublet scalar extension model are given as follows:
\begin{align}
c_{WW}&=c_{BB}=\frac{1}{(4\pi)^2}\frac{2\lambda_1+\lambda_2}{48}\frac{1}{M^2_\Phi}, \nn\\
c_{WB}&=\frac{1}{(4\pi)^2}\frac{\lambda_2}{24}\frac{1}{M^2_\Phi}, \nn\\
c_{2W}&=c_{3W}=\frac{1}{(4\pi)^2}\frac{g^2}{60}\frac{1}{M^2_\Phi},   \nn\\
c_{2B}&=\frac{1}{(4\pi)^2}\frac{g'^2}{60}\frac{1}{M^2_\Phi}, \nn\\
c_{T}&=\frac{1}{(4\pi)^2}\frac{\lambda_2^2-4\lambda_3^2}{12}\frac{1}{M^2_\Phi},  \nn\\
c_{r}&=\frac{1}{(4\pi)^2}\left(6\eta_\Phi\eta_H+\frac{\lambda_2^2+4\lambda_3^2}{6}
\right)\frac{1}{M^2_\Phi},\nn\\
c_{H}&=\frac{1}{(4\pi)^2}\left(6 \eta_\Phi \eta_H+\frac{4\lambda_1^2+4\lambda_1\lambda_2+\lambda_2^2+4\lambda_3^2}{12}\right)\frac{1}{M^2_\Phi}, \nn \\
c_6&=\eta_H^2+\frac{1}{(4\pi)^2}\left[\frac{3}{2}\lambda_\Phi \eta_H^2+6\eta_\Phi(\lambda_1+\lambda_2) \right. \nn\\
&\left.-\frac{2\lambda_1^3+3\lambda_1^2\lambda_2+3\lambda_1\lambda_2^2+\lambda_2^3}{6}-2(\lambda_1+\lambda_2)\lambda_3^2\right]\frac{1}{M^2_\Phi}.\nn\\
\label{tab:hid}
\end{align}
Similar to the triplet case, the terms with $1/(4\pi)^2$ factor are induced at one-loop order. Again, we require $\lambda_{1,2,3}<4\pi$ and $\lambda_\Phi<4\pi$ to ensure perturbativity and unitarity.

For simplicity, we let $\eta_H=\eta_{\Phi}=0$, which reduces the doublet scalar extension model to the so-called Inert Doublet Model~\cite{Ma:2006km,Barbieri:2006dq,Cao:2007rm}.
After running down to the EW scale, one obtains those non-zero coefficients of dim-6 operators relevant to the SFOPT condition, EW precision tests, and the discovery potential of CEPC. Again, we note that the contribution of fermionic operators can be safely ignored, and we focus on bosonic operators below.
Omitting the operators with negligible coefficients, we end up with a set of operators as
\be
\{\CO_{WW}, ~\CO_{BB}, ~\CO_{WB}, ~\CO_T, ~\CO_H, ~\CO_6\}.
\ee

For illustration, we choose $M_\Phi=1~{\rm TeV}$  and scan over the model parameter space of the doublet extension model. Figure~\ref{doublet_M} shows the parameter space that is compatible with SFOPT (green) and current EW and Higgs boson data (blue) in the plane of ($\lambda_1$, $\lambda_2$) with $\lambda_3=0$ (a) , ($\lambda_1$, $\lambda_3$) with $\lambda_2=0$ (b), and ($\lambda_2$, $\lambda_3$) with $\lambda_1=0$ (c).

Similar to the case of the triplet scalar model, the shape of the parameter regions in Fig.~\ref{doublet_M} can be understood by those Wilson coefficients obtained at the scale of $M_\Phi$ in Eq.~(\ref{tab:hid}).
For example, the oblique $T$-parameter imposes severe constraints on the coefficient of the $\CO_T$ operator, $c_T\sim(\lambda_2^2-4\lambda_3^2)$. Hence, the blue shaded regions and blue hatched regions are insensitive to $\lambda_1$ in Figs.~\ref{doublet_M}(a) and ~\ref{doublet_M}(b). However, the blue regions exhibit hyperbolic shapes in Fig.~\ref{doublet_M}(c). Again, the $T$-parameter constraint also validates the assumption of heavy $\Phi$ scalar decoupling we made above.
Note that three massive scalars emerge from the doublet $\Phi=[(\mathcal{H}^0+i\mathcal{A}^0)/\sqrt{2}, \mathcal{H}^-]$ after symmetry breaking, $H^T\to (0, v/\sqrt{2})$. The masses of the resulting new scalar particles are given by
\begin{eqnarray}
m^2(\mathcal{H}^\pm) &=& M_\Phi^2 + \lambda_1 v^2,\nn\\
m^2(\mathcal{H}^0) &=& M_\Phi^2 + (\lambda_1 + \lambda_2 + 2\lambda_3) v^2; \nn\\
m^2(\mathcal{A}^0) &=& M_\Phi^2 + (\lambda_1 + \lambda_2 - 2\lambda_3) v^2.
\end{eqnarray}
Vacuum stability demands~\cite{Barroso:2013awa}
\begin{eqnarray}
\lambda,\;\lambda_{\Phi}>0,\quad \lambda_1,\; \lambda_1+\lambda_2-\left|2\lambda_3\right|>-\sqrt{\lambda\lambda_\Phi},
\end{eqnarray}
where $\lambda$ is the Higgs quartic self coupling defined in Eq.~(\ref{v00}).
For large $\lambda_{1,2,3}$, one of the neutral scalars, $\mathcal{H}^0$ or $\mathcal{A}^0$, could be lighter than $M_\Phi$ and could potentially be around the EW scale. This would obviously violate the $\Phi$ decoupling assumption. In the parameter space allowed by vacuum stability and the $T$-parameter ($c_T\propto\lambda^2_2-4\lambda_3^2 \sim 10^{-4} ~\text{ TeV}^{-2}$), the masses of the three new scalars are larger than $M_\Phi$ as long as $\lambda_1\ge 0$,

The band of the SFOPT condition (green) is controlled by
\begin{eqnarray}
-c_6&=& \left(\lambda_1^3+(\lambda_1+\lambda_2)^3+12(\lambda_1+\lambda_2)\lambda_3^2\right)\frac{1}{6M_\Phi^2}.
\end{eqnarray}
In all three cases in Fig.~\ref{doublet_M}, there is enough parameter space for SFOPT after considering the current experimental data. From the cross section deviation of $Zh$ associated production (red dashed contours) and the region probed by EW precise measurements (blue dashed lines), we find that the CEPC is able to test this type of SFOPT.

\subsection{Model with weak singlet scalar}

\begin{figure*}
  \centering
    \includegraphics[scale=0.3]{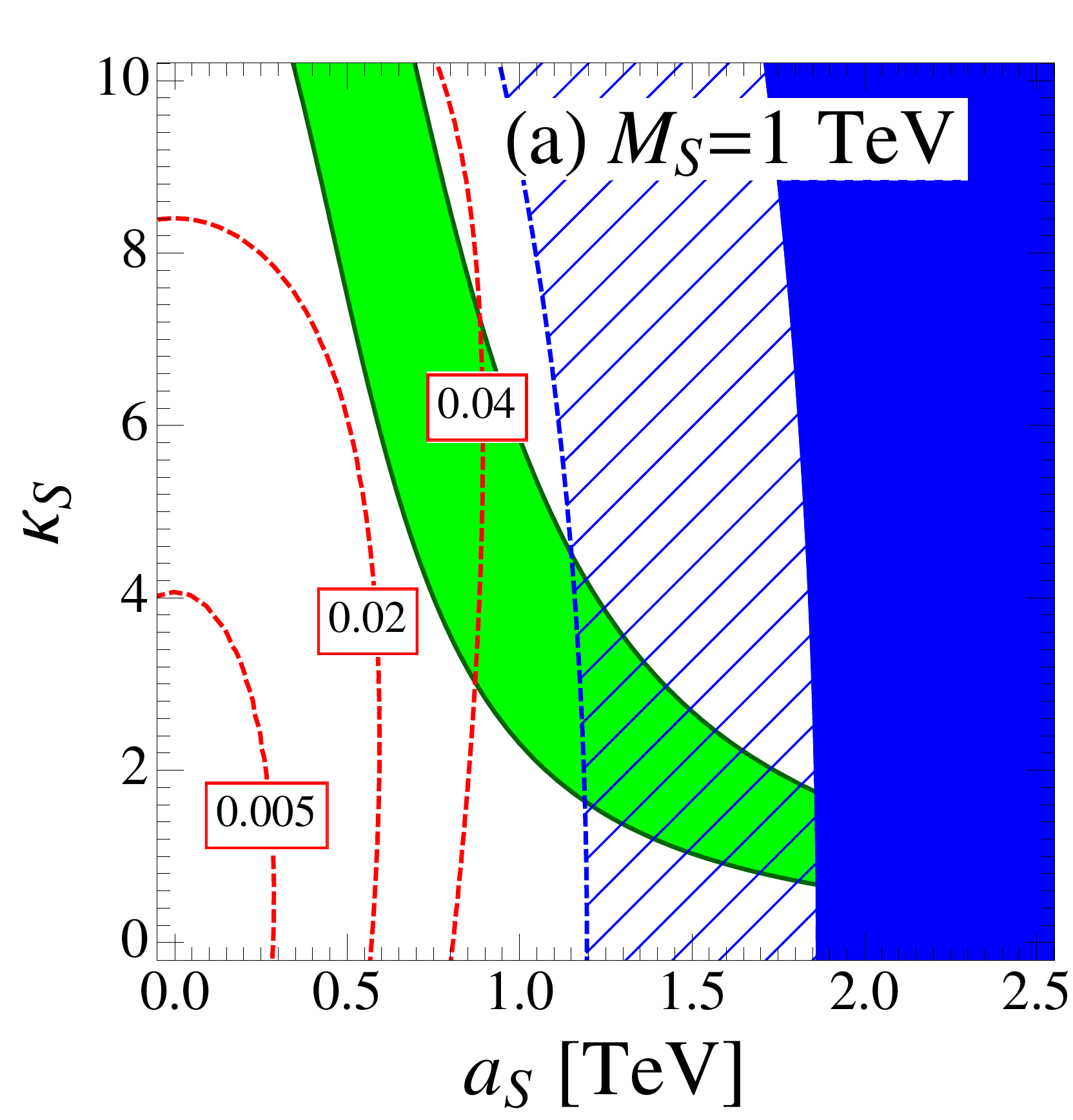}
    \includegraphics[scale=0.3]{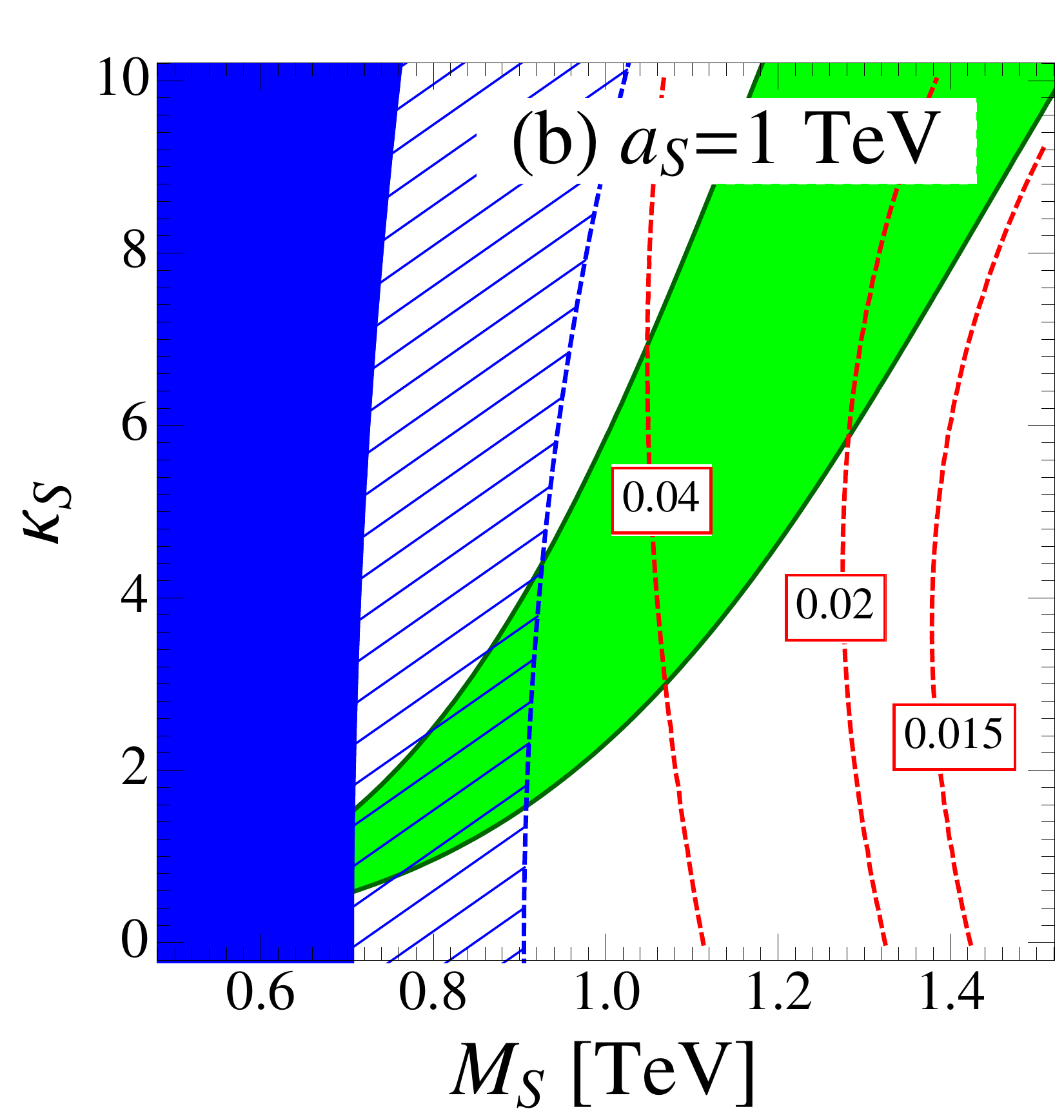}
    \includegraphics[scale=0.3]{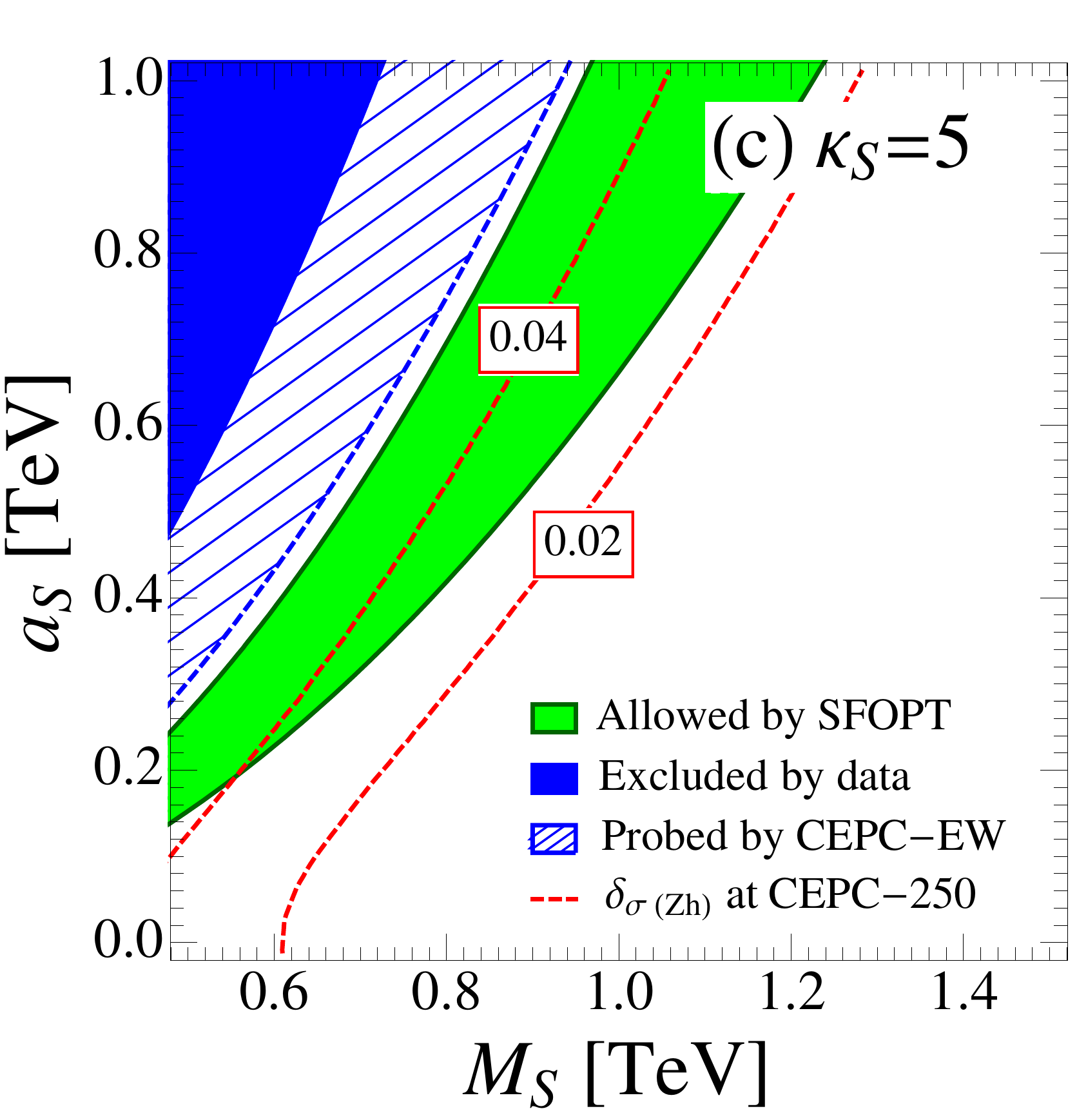}

\caption{The parameter space compatible with SFOPT (green), and current EWPT and Higgs boson data (blue) in the singlet scalar model. The region between the dashed blue line and the blue region is unconstrained by LEP results and can be explored by the CEPC. The red contours depict $\delta_{\sigma(Zh)}$ at  the CEPC.}
 \label{singlet_M}
\end{figure*}

Now consider the singlet scalar extension model~\cite{Chen:2017qcz}. We first investigate the case of a real scalar and then comment on the case of a complex scalar.

The relevant Lagrangian of a real singlet scalar~ is
\begin{eqnarray}\label{sl}
\delta\mathcal{L} &=&\frac{1}{2}\partial_\mu S\partial^\mu S-\frac{M_S^2}{2}S^2-\frac{\mu_S}{3!}S^3-\frac{\lambda_S}{4!} S^4\nn\\
&&-\frac{\kappa_S}{2}S^2 |H|^2-a_S S |H|^2.
\end{eqnarray}
Since the $S$ scalar is a gauge singlet, only two operators, $\CO_6$ and $\CO_H$, are generated at the $M_S$ scale. The corresponding dim-6 operators matched from the singlet scalar model up to next-to-leading order are
\begin{eqnarray}\label{seff}
\mathcal{L}_{\rm eff}&\supset&\Big(-\frac{\kappa_S a_S^2}{2 M_S^4}-\frac{1}{(4\pi)^2}\frac{\kappa_S^3}{12 M_S^2}+\frac{\mu_S a_S^3}{3!M_S^6}\Big)\CO_6\nn\\
&&+\Big(\frac{a_S^2}{M_S^4}+\frac{1}{(4\pi)^2}\frac{\kappa_S^2}{12 M_S^2}\Big)\CO_H.
\end{eqnarray}
Here, we consider $\kappa_S\lesssim10$ to make sure the model is perturbatively calculable, following Ref.~\cite{Curtin:2014jma}\footnote{In Ref.~\cite{Curtin:2014jma}, the perturbative condition is identified as $\lambda_{HS}\lesssim5$, and $\lambda_{HS}$ corresponds to $\kappa_S/2$ in our model.}. Owing to RG running effects, new operators such as $\CO_T$, $\CO_L^l$ and $\CO_R^e$ appear at the EW scale. However, their contributions to the EW precision observables and Higgs measurements are fairly small and can be safely ignored.
The corrections to the renormalizable operators in the Higgs potential can be absorbed into the redefinition of the coefficients of the
renormalizable operators, whose values are determined by the Higgs mass and the vacuum expectation value $v$.

The SFOPT condition is dominated by the leading term of the coefficient of $\CO_6$, i.e. $-c_6\sim \kappa_S a_S^2/2M_S^4$; see the green bands in Fig.~\ref{singlet_M}.
We emphasize that a non-zero $a_S$ in the real singlet scalar model is crucial to trigger SFOPT within the perturbative limit of $\kappa_S$. When the $a_S$ term is absent, the $\CO_6$ operator can be induced only at the one-loop level and thus suffers a large suppression from the factor of $1/(4\pi)^2$. In order to achieve the SFOPT (i.e. sizable $c_6$), a huge $\kappa_S$ is needed, e.g. $\kappa_S\sim 14$. Such a large coupling violates the perturbative limit mentioned above.

A non-zero $a_S$ term also yields a vacuum expectation value to the $S$ field,
\begin{eqnarray}
\langle S\rangle=\frac{-a_Sv^2/2}{M_S^2+\kappa_Sv^2/2},
\end{eqnarray}
after the Higgs field $H$ develops the vacuum expectation value $v$. The term also induces a mixing between the SM Higgs boson $h$ and $S$, which universally modifies the Higgs boson couplings to the SM particles such that it is strongly constrained by current LHC data. In the heavy $S$ limit, the sine of the mixing angle is $\sin\theta\approx va_S/M_S^2$. The ratio of the $hVV$ couplings in the real singlet scalar model and the SM is given by
\begin{eqnarray}
\kappa_V=\frac{g_{hVV}}{g_{hVV}^{\rm SM}}\approx1-\frac{v^2a_S^2}{2M_S^4}.
\end{eqnarray}
For the benchmark point $a_S=M_S=1$ TeV, $\kappa_V\sim0.97$, which is consistent with current experimental data~\cite{ATLAS-CONF-2015-044, ATLAS-CONF-2017-045}. We scan over the whole parameter space and find the parameter spaces of interest to us are all consistent with experimental data.

The EW precision observables are dominated by the operator $\CO_H$; note that $\CO_6$ is almost unconstrained at the LEP and LHC. As $\CO_H$ only modifies the Higgs boson interaction through the field redefinition of the Higgs field, it contributes a universal factor to the Higgs couplings without changing the Higgs-boson decay branching ratios. The low energy precision tests are nearly irrelevant to $\CO_H$, while the Higgs boson measurements at the LHC are not precise enough to constrain $\CO_H$. Therefore, most of the parameter space of the real singlet scalar model remains a blind region in current experimental searches and even at the CEPC in EW precision operation; see the solid blue  and hatched blue  regions in Fig.~\ref{singlet_M}. The parameter space can be widely probed at the 250 GeV CEPC through $Zh$ production; see the red contours. The shapes of $\delta_{\sigma(Zh)}$ are quite different from those of the SFOPT condition, as the former is determined by both $c_H$ and $c_6$ while the latter depends only on $c_6$.
In Ref.~\cite{Damgaard:2015con}, the authors claimed that the EFT approach for discussing the
phase transition may lead to some mismatches and miss some parameter spaces which are allowed in the renormalizable singlet model.
It is true that some allowed parameter spaces are missed in the EFT framework compared to the renormalizable model.
However, we are only concerned with whether there are allowed parameter spaces for this type of tree-level barrier SFOPT
induced by non-renormalizable operators (for a given renormalizable model, there may exist different types of SFOPT in different parameter spaces~\cite{{Chung:2012vg}}; we will not consider these aspects in this work), and the allowed parameter spaces are consistent with the results in Ref.~\cite{Damgaard:2015con}.

Let us proceed to the complex scalar extension model. The remarkable difference between the complex scalar and real scalar extension models is the former does not have the $a_S$ term while the latter does. The general Lagrangian of the singlet complex scalar $S$ is
\begin{eqnarray}\label{c_sl}
\delta\mathcal{L} &=&D_\mu S^\dagger D^\mu S-M_S^2|S|^2-\frac{\lambda_S}{4} |S|^4-\kappa_S|S|^2 |H|^2,~~~
\end{eqnarray}
where $D_\mu S=\partial_\mu S-ig'Y_SB_\mu S$, with $Y_S$ being the hypercharge of the singlet scalar $S$.
The $U(1)_Y$ symmetry of the complex scalar field forbids the linear and cubic terms of the complex scalar, i.e. no $a_S$ or $\mu_S$ terms as in the singlet real scalar model.
Using the CDE method, we obtain the dim-6 effective operators at the one-loop level as follows:
\begin{eqnarray}
\mathcal{L}_{\rm eff}&\supset&-\frac{1}{(4\pi)^2}\frac{\kappa_S^3}{6 M_S^2}\CO_6+\frac{1}{(4\pi)^2}\frac{\kappa_S^2}{6 M_S^2}\CO_H\nn\\
&&+\frac{1}{(4\pi)^2}\frac{\kappa_SY_S^2}{12M_S^2}\CO_{BB}+\frac{1}{(4\pi)^2}\frac{g'^2Y_S^2}{30M_S^2}\CO_{2B}.
\end{eqnarray}
The coefficients of $\CO_6$ and $\CO_H$ are twice those in the real scalar model (setting $a_S=\mu_S=0$), because a complex scalar is equivalence to two real scalars. Besides those pure scalar operators, we have two more operators involving gauge bosons, $\CO_{BB}$ and $\CO_{2B}$.

As discussed above, it is hard to explain the SFOPT in the perturbative parameter space in the complex scalar model. In the complex scalar model, the dim-6 operators are suppressed by a factor of $1/(4\pi)^2$. The parameter $\kappa_S$ has to be large enough to satisfy the SFOPT condition. In such a case, perturbativity may be violated; for example, for $M_S=1$ TeV, the SFOPT condition demands $\kappa_S\gtrsim10.7$, which is beyond the perturbative region. Of course, slightly reducing $M_S$ can render $\kappa_S$ marginally below the perturbative limit, e.g. $\kappa_S\gtrsim9.2$ for $M_S=0.8~{\rm TeV}$, but it is not natural. In addition, if $M_S$ is too small, the EFT description fails and one has to carry out the loop calculation in the UV-complete model. Therefore, we conclude that the $|H|^6$ SFOPT scenario is not favored in the singlet heavy complex-scalar model.

\subsection{General discussion}

From the above study, it is clear that the scenario of SFOPT induced by $|H|^6$ can still be realistic and allowed by current experimental data.
Our discussion on the dim-6 operators generated in the three scalar extension models
can be widely extended to many NP models.
Usually, the SFOPT needs the Higgs portal coupling to be of order one, and the large Higgs portal coupling may give a hint of the composite nature of the Higgs boson~\cite{Grinstein:2008qi}.
If the Higgs boson is a pseudo-Goldstone boson, from strong dynamics, the coefficients of dim-6 operators can be
estimated by naive dimensional analysis (NDA)~\cite{Georgi:1992dw}.
The coefficients of dominant CP-conserving operators, estimated from the NDA, are shown as follows:
\begin{align}
&c_{WW} \sim c_{BB} \sim c_{WB} \sim 1/\Lambda^2 \sim 1/(4\pi f)^2 \nn\\
&c_{H} \sim c_{T}\sim 1/f^2\nn\\
&c_6 \sim -\Lambda^2/f^4=-1/(f/4\pi)^2. \label{tab:hinaive}
\end{align}
If the EW phase transition is SFOPT, then one needs
\begin{equation*}
\frac{1}{(0.89 \rm~TeV)^2} < -c_6 <\frac{1}{(0.55~\rm TeV)^2},
\end{equation*}
which demands $$6.91~{\rm TeV}<f<11.18~{\rm TeV}.$$
The coefficients $c_{WW,BB,WB,H,T}$ are consistent with current experiments if the scale $f$ is within the above range.

The cross section deviation of $Zh$ production at the CEPC is approximately given by
\begin{eqnarray}
\label{ZH_Xsection}
\delta_{\sigma(Zh)} &\approx& (0.26 c_{WW} + 0.01 c_{BB} +0.04 c_{WB}-0.06 c_H\nn\\
&& -0.04 c_T)\times\text{ TeV}^2+0.016 \delta_h.
\end{eqnarray}
Without the $\delta_h$ term, we find $\delta_{\sigma(Zh)}\sim 0.1\%$ when choosing those coefficients shown in Eq.~(\ref{tab:hinaive}). The SFOPT condition requires $0.6 < \delta_h < 1.5$. Therefore, including the $\delta_h$ contribution yields $\delta_{\sigma(Zh)}$ in the range of (0.96 - 2.4)\%, which could be probed at future lepton colliders, such as the CEPC, ILC and FCC-ee.

\section{Conclusion}\label{sec:sum}
Unravelling the shape of the Higgs potential or the type of EW phase transition
is the central issue after the discovery of the Higgs boson.
The strong first-order EW phase transition provides a necessary condition for EW baryogenesis.
In this work we focus on the scenario that the strong first-order EW phase transition is induced by a $|H|^6$ operator. The operator can be generated by heavy particles or strong dynamics at some high scale (at which the excitations of the underlying theory can be directly probed). We have considered three new physics models with a scalar extension and examined the condition of strong first-order EW phase transitions in each new physics model. Specifically, the weak triplet, doublet and singlet complex/real scalar extension models have been considered here. The three models represent the structure of the scalar sector of many new physics models. The $|H|^6$ operators are induced when those new scalars are decoupled at some heavy scale. Simultaneously, many other dim-6 operators will also be generated and their coefficients are highly correlated with the coefficient of the $|H|^6$ operator, because the coefficients depend on the same set of model parameters in a given new physics model. While making use of $|H|^6$ to generate the EW strong first-order phase transition, one has to check whether the model parameters have been ruled out by EW precision tests and other experiments such as Higgs boson physics.

We have used the CDE method to derive all the dimension-6 effective operators in the three scalar extension models. We have found that the triplet scalar, doublet scalar and singlet real scalar extension models can generate the EW SFOPT without violating the perturbative limit. However, it is hard to address the EW SFOPT in the singlet complex-scalar extension model when the singlet scalars are very heavy.  We have performed a global fit to provide bounds on the complete set of dimension-6 operators by including the EW precision test and recent Higgs measurements. We have found that the parameter space of SFOPT can be probed extensively in $Zh$ production at future electron-positron colliders, such as the CEPC, ILC and FCC-ee.

\section*{Acknowledgments}
We thank Adam Falkowski, Shao-Feng Ge, Hitoshi Murayama, Francesco Riva, Lian-Tao Wang and Hao Zhang for useful discussions.

QHC and KPX are supported in part by the National Science Foundation of China under Grand No. 11175069, No. 11275009 and No. 11422545.
XZ and FPH are supported by the NSFC under grants Nos. 11121092, 11033005, 11375202
and also by the CAS Pilot-B program. FPH is also supported by the China Postdoctoral Science Foundation under Grant Nos. 2016M590133 and 2017T100108.
\bibliographystyle{apsrev}
\bibliography{reference}

\begin{thebibliography}{94}
\expandafter\ifx\csname natexlab\endcsname\relax\def\natexlab#1{#1}\fi
\expandafter\ifx\csname bibnamefont\endcsname\relax
  \def\bibnamefont#1{#1}\fi
\expandafter\ifx\csname bibfnamefont\endcsname\relax
  \def\bibfnamefont#1{#1}\fi
\expandafter\ifx\csname citenamefont\endcsname\relax
  \def\citenamefont#1{#1}\fi
\expandafter\ifx\csname url\endcsname\relax
  \def\url#1{\texttt{#1}}\fi
\expandafter\ifx\csname urlprefix\endcsname\relax\def\urlprefix{URL }\fi
\providecommand{\bibinfo}[2]{#2}
\providecommand{\eprint}[2][]{\url{#2}}

\bibitem[{\citenamefont{Aad et~al.}(2012)}]{Aad:2012tfa}
\bibinfo{author}{\bibfnamefont{G.}~\bibnamefont{Aad}} \bibnamefont{et~al.}
  (\bibinfo{collaboration}{ATLAS}), \bibinfo{journal}{Phys. Lett.}
  \textbf{\bibinfo{volume}{B716}}, \bibinfo{pages}{1} (\bibinfo{year}{2012}),
  \eprint{1207.7214}.

\bibitem[{\citenamefont{Chatrchyan et~al.}(2012)}]{Chatrchyan:2012xdj}
\bibinfo{author}{\bibfnamefont{S.}~\bibnamefont{Chatrchyan}}
  \bibnamefont{et~al.} (\bibinfo{collaboration}{CMS}), \bibinfo{journal}{Phys.
  Lett.} \textbf{\bibinfo{volume}{B716}}, \bibinfo{pages}{30}
  (\bibinfo{year}{2012}), \eprint{1207.7235}.

\bibitem[{\citenamefont{Ade et~al.}(2014)}]{Ade:2013zuv}
\bibinfo{author}{\bibfnamefont{P.~A.~R.} \bibnamefont{Ade}}
  \bibnamefont{et~al.} (\bibinfo{collaboration}{Planck}),
  \bibinfo{journal}{Astron. Astrophys.} \textbf{\bibinfo{volume}{571}},
  \bibinfo{pages}{A16} (\bibinfo{year}{2014}), \eprint{1303.5076}.

\bibitem[{\citenamefont{Patrignani et~al.}(2016)}]{Olive:2016xmw}
\bibinfo{author}{\bibfnamefont{C.}~\bibnamefont{Patrignani}}
  \bibnamefont{et~al.} (\bibinfo{collaboration}{Particle Data Group}),
  \bibinfo{journal}{Chin. Phys.} \textbf{\bibinfo{volume}{C40}},
  \bibinfo{pages}{100001} (\bibinfo{year}{2016}).

\bibitem[{\citenamefont{Dine and Kusenko}(2003)}]{Dine:2003ax}
\bibinfo{author}{\bibfnamefont{M.}~\bibnamefont{Dine}} \bibnamefont{and}
  \bibinfo{author}{\bibfnamefont{A.}~\bibnamefont{Kusenko}},
  \bibinfo{journal}{Rev. Mod. Phys.} \textbf{\bibinfo{volume}{76}},
  \bibinfo{pages}{1} (\bibinfo{year}{2003}), \eprint{hep-ph/0303065}.

\bibitem[{\citenamefont{Sakharov}(1967)}]{Sakharov:1967dj}
\bibinfo{author}{\bibfnamefont{A.~D.} \bibnamefont{Sakharov}},
  \bibinfo{journal}{Pisma Zh. Eksp. Teor. Fiz.} \textbf{\bibinfo{volume}{5}},
  \bibinfo{pages}{32} (\bibinfo{year}{1967}), \bibinfo{note}{[Usp. Fiz.
  Nauk161,61(1991)]}.

\bibitem[{\citenamefont{Kuzmin et~al.}(1985)\citenamefont{Kuzmin, Rubakov, and
  Shaposhnikov}}]{Kuzmin:1985mm}
\bibinfo{author}{\bibfnamefont{V.~A.} \bibnamefont{Kuzmin}},
  \bibinfo{author}{\bibfnamefont{V.~A.} \bibnamefont{Rubakov}},
  \bibnamefont{and} \bibinfo{author}{\bibfnamefont{M.~E.}
  \bibnamefont{Shaposhnikov}}, \bibinfo{journal}{Phys. Lett.}
  \textbf{\bibinfo{volume}{155B}}, \bibinfo{pages}{36} (\bibinfo{year}{1985}).

\bibitem[{\citenamefont{Trodden}(1999)}]{Trodden:1998ym}
\bibinfo{author}{\bibfnamefont{M.}~\bibnamefont{Trodden}},
  \bibinfo{journal}{Rev. Mod. Phys.} \textbf{\bibinfo{volume}{71}},
  \bibinfo{pages}{1463} (\bibinfo{year}{1999}), \eprint{hep-ph/9803479}.

\bibitem[{\citenamefont{Morrissey and Ramsey-Musolf}(2012)}]{Morrissey:2012db}
\bibinfo{author}{\bibfnamefont{D.~E.} \bibnamefont{Morrissey}}
  \bibnamefont{and} \bibinfo{author}{\bibfnamefont{M.~J.}
  \bibnamefont{Ramsey-Musolf}}, \bibinfo{journal}{New J. Phys.}
  \textbf{\bibinfo{volume}{14}}, \bibinfo{pages}{125003}
  (\bibinfo{year}{2012}), \eprint{1206.2942}.

\bibitem[{\citenamefont{Witten}(1984)}]{Witten:1984rs}
\bibinfo{author}{\bibfnamefont{E.}~\bibnamefont{Witten}},
  \bibinfo{journal}{Phys. Rev.} \textbf{\bibinfo{volume}{D30}},
  \bibinfo{pages}{272} (\bibinfo{year}{1984}).

\bibitem[{\citenamefont{Hogan}(1983)}]{Hogan:1984hx}
\bibinfo{author}{\bibfnamefont{C.~J.} \bibnamefont{Hogan}},
  \bibinfo{journal}{Phys. Lett.} \textbf{\bibinfo{volume}{133B}},
  \bibinfo{pages}{172} (\bibinfo{year}{1983}).

\bibitem[{\citenamefont{Turner and Wilczek}(1990)}]{Turner:1990rc}
\bibinfo{author}{\bibfnamefont{M.~S.} \bibnamefont{Turner}} \bibnamefont{and}
  \bibinfo{author}{\bibfnamefont{F.}~\bibnamefont{Wilczek}},
  \bibinfo{journal}{Phys. Rev. Lett.} \textbf{\bibinfo{volume}{65}},
  \bibinfo{pages}{3080} (\bibinfo{year}{1990}).

\bibitem[{\citenamefont{Kamionkowski et~al.}(1994)\citenamefont{Kamionkowski,
  Kosowsky, and Turner}}]{Kamionkowski:1993fg}
\bibinfo{author}{\bibfnamefont{M.}~\bibnamefont{Kamionkowski}},
  \bibinfo{author}{\bibfnamefont{A.}~\bibnamefont{Kosowsky}}, \bibnamefont{and}
  \bibinfo{author}{\bibfnamefont{M.~S.} \bibnamefont{Turner}},
  \bibinfo{journal}{Phys. Rev.} \textbf{\bibinfo{volume}{D49}},
  \bibinfo{pages}{2837} (\bibinfo{year}{1994}), \eprint{astro-ph/9310044}.

\bibitem[{\citenamefont{Kosowsky et~al.}(2002)\citenamefont{Kosowsky, Mack, and
  Kahniashvili}}]{Kosowsky:2001xp}
\bibinfo{author}{\bibfnamefont{A.}~\bibnamefont{Kosowsky}},
  \bibinfo{author}{\bibfnamefont{A.}~\bibnamefont{Mack}}, \bibnamefont{and}
  \bibinfo{author}{\bibfnamefont{T.}~\bibnamefont{Kahniashvili}},
  \bibinfo{journal}{Phys. Rev.} \textbf{\bibinfo{volume}{D66}},
  \bibinfo{pages}{024030} (\bibinfo{year}{2002}), \eprint{astro-ph/0111483}.

\bibitem[{\citenamefont{Caprini et~al.}(2009)\citenamefont{Caprini, Durrer, and
  Servant}}]{Caprini:2009yp}
\bibinfo{author}{\bibfnamefont{C.}~\bibnamefont{Caprini}},
  \bibinfo{author}{\bibfnamefont{R.}~\bibnamefont{Durrer}}, \bibnamefont{and}
  \bibinfo{author}{\bibfnamefont{G.}~\bibnamefont{Servant}},
  \bibinfo{journal}{JCAP} \textbf{\bibinfo{volume}{0912}}, \bibinfo{pages}{024}
  (\bibinfo{year}{2009}), \eprint{0909.0622}.

\bibitem[{\citenamefont{Hindmarsh et~al.}(2014)\citenamefont{Hindmarsh, Huber,
  Rummukainen, and Weir}}]{Hindmarsh:2013xza}
\bibinfo{author}{\bibfnamefont{M.}~\bibnamefont{Hindmarsh}},
  \bibinfo{author}{\bibfnamefont{S.~J.} \bibnamefont{Huber}},
  \bibinfo{author}{\bibfnamefont{K.}~\bibnamefont{Rummukainen}},
  \bibnamefont{and} \bibinfo{author}{\bibfnamefont{D.~J.} \bibnamefont{Weir}},
  \bibinfo{journal}{Phys. Rev. Lett.} \textbf{\bibinfo{volume}{112}},
  \bibinfo{pages}{041301} (\bibinfo{year}{2014}), \eprint{1304.2433}.

\bibitem[{\citenamefont{Hindmarsh et~al.}(2015)\citenamefont{Hindmarsh, Huber,
  Rummukainen, and Weir}}]{Hindmarsh:2015qta}
\bibinfo{author}{\bibfnamefont{M.}~\bibnamefont{Hindmarsh}},
  \bibinfo{author}{\bibfnamefont{S.~J.} \bibnamefont{Huber}},
  \bibinfo{author}{\bibfnamefont{K.}~\bibnamefont{Rummukainen}},
  \bibnamefont{and} \bibinfo{author}{\bibfnamefont{D.~J.} \bibnamefont{Weir}},
  \bibinfo{journal}{Phys. Rev.} \textbf{\bibinfo{volume}{D92}},
  \bibinfo{pages}{123009} (\bibinfo{year}{2015}), \eprint{1504.03291}.

\bibitem[{\citenamefont{Chung et~al.}(2013)\citenamefont{Chung, Long, and
  Wang}}]{Chung:2012vg}
\bibinfo{author}{\bibfnamefont{D.~J.~H.} \bibnamefont{Chung}},
  \bibinfo{author}{\bibfnamefont{A.~J.} \bibnamefont{Long}}, \bibnamefont{and}
  \bibinfo{author}{\bibfnamefont{L.-T.} \bibnamefont{Wang}},
  \bibinfo{journal}{Phys. Rev.} \textbf{\bibinfo{volume}{D87}},
  \bibinfo{pages}{023509} (\bibinfo{year}{2013}), \eprint{1209.1819}.

\bibitem[{\citenamefont{Grojean et~al.}(2005)\citenamefont{Grojean, Servant,
  and Wells}}]{Grojean:2004xa}
\bibinfo{author}{\bibfnamefont{C.}~\bibnamefont{Grojean}},
  \bibinfo{author}{\bibfnamefont{G.}~\bibnamefont{Servant}}, \bibnamefont{and}
  \bibinfo{author}{\bibfnamefont{J.~D.} \bibnamefont{Wells}},
  \bibinfo{journal}{Phys. Rev.} \textbf{\bibinfo{volume}{D71}},
  \bibinfo{pages}{036001} (\bibinfo{year}{2005}), \eprint{hep-ph/0407019}.

\bibitem[{\citenamefont{Zhang}(1993)}]{Zhang:1992fs}
\bibinfo{author}{\bibfnamefont{X.-m.} \bibnamefont{Zhang}},
  \bibinfo{journal}{Phys. Rev.} \textbf{\bibinfo{volume}{D47}},
  \bibinfo{pages}{3065} (\bibinfo{year}{1993}), \eprint{hep-ph/9301277}.

\bibitem[{\citenamefont{Zhang and Young}(1994)}]{Zhang:1993vh}
\bibinfo{author}{\bibfnamefont{X.}~\bibnamefont{Zhang}} \bibnamefont{and}
  \bibinfo{author}{\bibfnamefont{B.~L.} \bibnamefont{Young}},
  \bibinfo{journal}{Phys. Rev.} \textbf{\bibinfo{volume}{D49}},
  \bibinfo{pages}{563} (\bibinfo{year}{1994}), \eprint{hep-ph/9309269}.

\bibitem[{\citenamefont{Whisnant et~al.}(1995)\citenamefont{Whisnant, Young,
  and Zhang}}]{Whisnant:1994fh}
\bibinfo{author}{\bibfnamefont{K.}~\bibnamefont{Whisnant}},
  \bibinfo{author}{\bibfnamefont{B.-L.} \bibnamefont{Young}}, \bibnamefont{and}
  \bibinfo{author}{\bibfnamefont{X.}~\bibnamefont{Zhang}},
  \bibinfo{journal}{Phys. Rev.} \textbf{\bibinfo{volume}{D52}},
  \bibinfo{pages}{3115} (\bibinfo{year}{1995}), \eprint{hep-ph/9410369}.

\bibitem[{\citenamefont{Zhang et~al.}(1994)\citenamefont{Zhang, Lee, Whisnant,
  and Young}}]{Zhang:1994fb}
\bibinfo{author}{\bibfnamefont{X.}~\bibnamefont{Zhang}},
  \bibinfo{author}{\bibfnamefont{S.~K.} \bibnamefont{Lee}},
  \bibinfo{author}{\bibfnamefont{K.}~\bibnamefont{Whisnant}}, \bibnamefont{and}
  \bibinfo{author}{\bibfnamefont{B.~L.} \bibnamefont{Young}},
  \bibinfo{journal}{Phys. Rev.} \textbf{\bibinfo{volume}{D50}},
  \bibinfo{pages}{7042} (\bibinfo{year}{1994}), \eprint{hep-ph/9407259}.

\bibitem[{\citenamefont{Huang and Li}(2015)}]{Huang:2015bta}
\bibinfo{author}{\bibfnamefont{F.~P.} \bibnamefont{Huang}} \bibnamefont{and}
  \bibinfo{author}{\bibfnamefont{C.~S.} \bibnamefont{Li}},
  \bibinfo{journal}{Phys. Rev.} \textbf{\bibinfo{volume}{D92}},
  \bibinfo{pages}{075014} (\bibinfo{year}{2015}), \eprint{1507.08168}.

\bibitem[{\citenamefont{Kobakhidze et~al.}(2016)\citenamefont{Kobakhidze, Wu,
  and Yue}}]{Kobakhidze:2015xlz}
\bibinfo{author}{\bibfnamefont{A.}~\bibnamefont{Kobakhidze}},
  \bibinfo{author}{\bibfnamefont{L.}~\bibnamefont{Wu}}, \bibnamefont{and}
  \bibinfo{author}{\bibfnamefont{J.}~\bibnamefont{Yue}},
  \bibinfo{journal}{JHEP} \textbf{\bibinfo{volume}{04}}, \bibinfo{pages}{011}
  (\bibinfo{year}{2016}), \eprint{1512.08922}.

\bibitem[{\citenamefont{Balazs et~al.}(2017)\citenamefont{Balazs, White, and
  Yue}}]{Balazs:2016yvi}
\bibinfo{author}{\bibfnamefont{C.}~\bibnamefont{Balazs}},
  \bibinfo{author}{\bibfnamefont{G.}~\bibnamefont{White}}, \bibnamefont{and}
  \bibinfo{author}{\bibfnamefont{J.}~\bibnamefont{Yue}},
  \bibinfo{journal}{JHEP} \textbf{\bibinfo{volume}{03}}, \bibinfo{pages}{030}
  (\bibinfo{year}{2017}), \eprint{1612.01270}.

\bibitem[{\citenamefont{Henning et~al.}(2016)\citenamefont{Henning, Lu, and
  Murayama}}]{Henning:2014wua}
\bibinfo{author}{\bibfnamefont{B.}~\bibnamefont{Henning}},
  \bibinfo{author}{\bibfnamefont{X.}~\bibnamefont{Lu}}, \bibnamefont{and}
  \bibinfo{author}{\bibfnamefont{H.}~\bibnamefont{Murayama}},
  \bibinfo{journal}{JHEP} \textbf{\bibinfo{volume}{01}}, \bibinfo{pages}{023}
  (\bibinfo{year}{2016}), \eprint{1412.1837}.

\bibitem[{CEP(2015)}]{CEPC-SPPCStudyGroup:2015csa}
\bibinfo{type}{Tech. Rep.} \bibinfo{number}{IHEP-CEPC-DR-2015-01},
  \bibinfo{institution}{CEPC-SPPC Study Group} (\bibinfo{year}{2015}).

\bibitem[{\citenamefont{Bicer et~al.}(2014)}]{Gomez-Ceballos:2013zzn}
\bibinfo{author}{\bibfnamefont{M.}~\bibnamefont{Bicer}} \bibnamefont{et~al.}
  (\bibinfo{collaboration}{TLEP Design Study Working Group}),
  \bibinfo{journal}{JHEP} \textbf{\bibinfo{volume}{01}}, \bibinfo{pages}{164}
  (\bibinfo{year}{2014}), \eprint{1308.6176}.

\bibitem[{\citenamefont{d'Enterria}(2016{\natexlab{a}})}]{dEnterria:2016fpc}
\bibinfo{author}{\bibfnamefont{D.}~\bibnamefont{d'Enterria}},
  \bibinfo{journal}{Frascati Phys. Ser.} \textbf{\bibinfo{volume}{61}},
  \bibinfo{pages}{17} (\bibinfo{year}{2016}{\natexlab{a}}),
  \eprint{1601.06640}.

\bibitem[{\citenamefont{d'Enterria}(2016{\natexlab{b}})}]{d'Enterria:2132590}
\bibinfo{author}{\bibfnamefont{D.}~\bibnamefont{d'Enterria}},
  \bibinfo{type}{Tech. Rep.} \bibinfo{number}{arXiv:1602.05043}
  (\bibinfo{year}{2016}{\natexlab{b}}), \bibinfo{note}{comments: 8 pages, 8
  figures. Proceedings 17th Lomonosov conference on Elementary Particle
  Physics, Moscow, Aug. 2015. World Scientific, to appear. arXiv admin note:
  text overlap with arXiv:1601.06640},
  \urlprefix\url{http://cds.cern.ch/record/2132590}.

\bibitem[{\citenamefont{Georgi}(1991)}]{Georgi:1991ch}
\bibinfo{author}{\bibfnamefont{H.}~\bibnamefont{Georgi}},
  \bibinfo{journal}{Nucl. Phys.} \textbf{\bibinfo{volume}{B361}},
  \bibinfo{pages}{339} (\bibinfo{year}{1991}).

\bibitem[{\citenamefont{Weinberg}(1979{\natexlab{a}})}]{Weinberg:1978kz}
\bibinfo{author}{\bibfnamefont{S.}~\bibnamefont{Weinberg}},
  \bibinfo{journal}{Physica} \textbf{\bibinfo{volume}{A96}},
  \bibinfo{pages}{327} (\bibinfo{year}{1979}{\natexlab{a}}).

\bibitem[{\citenamefont{Wudka}(1994)}]{Wudka:1994ny}
\bibinfo{author}{\bibfnamefont{J.}~\bibnamefont{Wudka}}, \bibinfo{journal}{Int.
  J. Mod. Phys.} \textbf{\bibinfo{volume}{A9}}, \bibinfo{pages}{2301}
  (\bibinfo{year}{1994}), \eprint{hep-ph/9406205}.

\bibitem[{\citenamefont{Weinberg}(1979{\natexlab{b}})}]{Weinberg:1979sa}
\bibinfo{author}{\bibfnamefont{S.}~\bibnamefont{Weinberg}},
  \bibinfo{journal}{Phys. Rev. Lett.} \textbf{\bibinfo{volume}{43}},
  \bibinfo{pages}{1566} (\bibinfo{year}{1979}{\natexlab{b}}).

\bibitem[{\citenamefont{Wilczek and Zee}(1979)}]{Wilczek:1979hc}
\bibinfo{author}{\bibfnamefont{F.}~\bibnamefont{Wilczek}} \bibnamefont{and}
  \bibinfo{author}{\bibfnamefont{A.}~\bibnamefont{Zee}},
  \bibinfo{journal}{Phys. Rev. Lett.} \textbf{\bibinfo{volume}{43}},
  \bibinfo{pages}{1571} (\bibinfo{year}{1979}).

\bibitem[{\citenamefont{Weldon and Zee}(1980)}]{Weldon:1980gi}
\bibinfo{author}{\bibfnamefont{H.~A.} \bibnamefont{Weldon}} \bibnamefont{and}
  \bibinfo{author}{\bibfnamefont{A.}~\bibnamefont{Zee}},
  \bibinfo{journal}{Nucl. Phys.} \textbf{\bibinfo{volume}{B173}},
  \bibinfo{pages}{269} (\bibinfo{year}{1980}).

\bibitem[{\citenamefont{Buchmuller and Wyler}(1986)}]{Buchmuller:1985jz}
\bibinfo{author}{\bibfnamefont{W.}~\bibnamefont{Buchmuller}} \bibnamefont{and}
  \bibinfo{author}{\bibfnamefont{D.}~\bibnamefont{Wyler}},
  \bibinfo{journal}{Nucl. Phys.} \textbf{\bibinfo{volume}{B268}},
  \bibinfo{pages}{621} (\bibinfo{year}{1986}).

\bibitem[{\citenamefont{Falkowski and Riva}(2015)}]{Falkowski:2014tna}
\bibinfo{author}{\bibfnamefont{A.}~\bibnamefont{Falkowski}} \bibnamefont{and}
  \bibinfo{author}{\bibfnamefont{F.}~\bibnamefont{Riva}},
  \bibinfo{journal}{JHEP} \textbf{\bibinfo{volume}{02}}, \bibinfo{pages}{039}
  (\bibinfo{year}{2015}), \eprint{1411.0669}.

\bibitem[{\citenamefont{Elias-Miro et~al.}(2013)\citenamefont{Elias-Miro,
  Espinosa, Masso, and Pomarol}}]{Elias-Miro:2013mua}
\bibinfo{author}{\bibfnamefont{J.}~\bibnamefont{Elias-Miro}},
  \bibinfo{author}{\bibfnamefont{J.~R.} \bibnamefont{Espinosa}},
  \bibinfo{author}{\bibfnamefont{E.}~\bibnamefont{Masso}}, \bibnamefont{and}
  \bibinfo{author}{\bibfnamefont{A.}~\bibnamefont{Pomarol}},
  \bibinfo{journal}{JHEP} \textbf{\bibinfo{volume}{11}}, \bibinfo{pages}{066}
  (\bibinfo{year}{2013}), \eprint{1308.1879}.

\bibitem[{\citenamefont{Alonso et~al.}(2014)\citenamefont{Alonso, Jenkins,
  Manohar, and Trott}}]{Alonso:2013hga}
\bibinfo{author}{\bibfnamefont{R.}~\bibnamefont{Alonso}},
  \bibinfo{author}{\bibfnamefont{E.~E.} \bibnamefont{Jenkins}},
  \bibinfo{author}{\bibfnamefont{A.~V.} \bibnamefont{Manohar}},
  \bibnamefont{and} \bibinfo{author}{\bibfnamefont{M.}~\bibnamefont{Trott}},
  \bibinfo{journal}{JHEP} \textbf{\bibinfo{volume}{04}}, \bibinfo{pages}{159}
  (\bibinfo{year}{2014}), \eprint{1312.2014}.

\bibitem[{\citenamefont{Brivio et~al.}(2017)\citenamefont{Brivio, Jiang, and
  Trott}}]{Brivio:2017btx}
\bibinfo{author}{\bibfnamefont{I.}~\bibnamefont{Brivio}},
  \bibinfo{author}{\bibfnamefont{Y.}~\bibnamefont{Jiang}}, \bibnamefont{and}
  \bibinfo{author}{\bibfnamefont{M.}~\bibnamefont{Trott}}
  (\bibinfo{year}{2017}), \eprint{1709.06492}.

\bibitem[{\citenamefont{Grzadkowski et~al.}(2010)\citenamefont{Grzadkowski,
  Iskrzynski, Misiak, and Rosiek}}]{Grzadkowski:2010es}
\bibinfo{author}{\bibfnamefont{B.}~\bibnamefont{Grzadkowski}},
  \bibinfo{author}{\bibfnamefont{M.}~\bibnamefont{Iskrzynski}},
  \bibinfo{author}{\bibfnamefont{M.}~\bibnamefont{Misiak}}, \bibnamefont{and}
  \bibinfo{author}{\bibfnamefont{J.}~\bibnamefont{Rosiek}},
  \bibinfo{journal}{JHEP} \textbf{\bibinfo{volume}{10}}, \bibinfo{pages}{085}
  (\bibinfo{year}{2010}), \eprint{1008.4884}.

\bibitem[{\citenamefont{Hagiwara et~al.}(1993)\citenamefont{Hagiwara, Ishihara,
  Szalapski, and Zeppenfeld}}]{Hagiwara:1993ck}
\bibinfo{author}{\bibfnamefont{K.}~\bibnamefont{Hagiwara}},
  \bibinfo{author}{\bibfnamefont{S.}~\bibnamefont{Ishihara}},
  \bibinfo{author}{\bibfnamefont{R.}~\bibnamefont{Szalapski}},
  \bibnamefont{and}
  \bibinfo{author}{\bibfnamefont{D.}~\bibnamefont{Zeppenfeld}},
  \bibinfo{journal}{Phys. Rev.} \textbf{\bibinfo{volume}{D48}},
  \bibinfo{pages}{2182} (\bibinfo{year}{1993}).

\bibitem[{\citenamefont{Giudice et~al.}(2007)\citenamefont{Giudice, Grojean,
  Pomarol, and Rattazzi}}]{Giudice:2007fh}
\bibinfo{author}{\bibfnamefont{G.~F.} \bibnamefont{Giudice}},
  \bibinfo{author}{\bibfnamefont{C.}~\bibnamefont{Grojean}},
  \bibinfo{author}{\bibfnamefont{A.}~\bibnamefont{Pomarol}}, \bibnamefont{and}
  \bibinfo{author}{\bibfnamefont{R.}~\bibnamefont{Rattazzi}},
  \bibinfo{journal}{JHEP} \textbf{\bibinfo{volume}{06}}, \bibinfo{pages}{045}
  (\bibinfo{year}{2007}), \eprint{hep-ph/0703164}.

\bibitem[{\citenamefont{Pomarol and Riva}(2014)}]{Pomarol:2013zra}
\bibinfo{author}{\bibfnamefont{A.}~\bibnamefont{Pomarol}} \bibnamefont{and}
  \bibinfo{author}{\bibfnamefont{F.}~\bibnamefont{Riva}},
  \bibinfo{journal}{JHEP} \textbf{\bibinfo{volume}{01}}, \bibinfo{pages}{151}
  (\bibinfo{year}{2014}), \eprint{1308.2803}.

\bibitem[{\citenamefont{Elias-Miro et~al.}(2014)\citenamefont{Elias-Miro,
  Grojean, Gupta, and Marzocca}}]{Elias-Miro:2013eta}
\bibinfo{author}{\bibfnamefont{J.}~\bibnamefont{Elias-Miro}},
  \bibinfo{author}{\bibfnamefont{C.}~\bibnamefont{Grojean}},
  \bibinfo{author}{\bibfnamefont{R.~S.} \bibnamefont{Gupta}}, \bibnamefont{and}
  \bibinfo{author}{\bibfnamefont{D.}~\bibnamefont{Marzocca}},
  \bibinfo{journal}{JHEP} \textbf{\bibinfo{volume}{05}}, \bibinfo{pages}{019}
  (\bibinfo{year}{2014}), \eprint{1312.2928}.

\bibitem[{\citenamefont{Delaunay et~al.}(2008)\citenamefont{Delaunay, Grojean,
  and Wells}}]{Delaunay:2007wb}
\bibinfo{author}{\bibfnamefont{C.}~\bibnamefont{Delaunay}},
  \bibinfo{author}{\bibfnamefont{C.}~\bibnamefont{Grojean}}, \bibnamefont{and}
  \bibinfo{author}{\bibfnamefont{J.~D.} \bibnamefont{Wells}},
  \bibinfo{journal}{JHEP} \textbf{\bibinfo{volume}{04}}, \bibinfo{pages}{029}
  (\bibinfo{year}{2008}), \eprint{0711.2511}.

\bibitem[{\citenamefont{Grinstein and Trott}(2008)}]{Grinstein:2008qi}
\bibinfo{author}{\bibfnamefont{B.}~\bibnamefont{Grinstein}} \bibnamefont{and}
  \bibinfo{author}{\bibfnamefont{M.}~\bibnamefont{Trott}},
  \bibinfo{journal}{Phys. Rev.} \textbf{\bibinfo{volume}{D78}},
  \bibinfo{pages}{075022} (\bibinfo{year}{2008}), \eprint{0806.1971}.

\bibitem[{\citenamefont{Ham and Oh}(2004)}]{Ham:2004zs}
\bibinfo{author}{\bibfnamefont{S.~W.} \bibnamefont{Ham}} \bibnamefont{and}
  \bibinfo{author}{\bibfnamefont{S.~K.} \bibnamefont{Oh}},
  \bibinfo{journal}{Phys. Rev.} \textbf{\bibinfo{volume}{D70}},
  \bibinfo{pages}{093007} (\bibinfo{year}{2004}), \eprint{hep-ph/0408324}.

\bibitem[{\citenamefont{Bodeker et~al.}(2005)\citenamefont{Bodeker, Fromme,
  Huber, and Seniuch}}]{Bodeker:2004ws}
\bibinfo{author}{\bibfnamefont{D.}~\bibnamefont{Bodeker}},
  \bibinfo{author}{\bibfnamefont{L.}~\bibnamefont{Fromme}},
  \bibinfo{author}{\bibfnamefont{S.~J.} \bibnamefont{Huber}}, \bibnamefont{and}
  \bibinfo{author}{\bibfnamefont{M.}~\bibnamefont{Seniuch}},
  \bibinfo{journal}{JHEP} \textbf{\bibinfo{volume}{02}}, \bibinfo{pages}{026}
  (\bibinfo{year}{2005}), \eprint{hep-ph/0412366}.

\bibitem[{\citenamefont{Chu et~al.}(2015)\citenamefont{Chu, Jansen,
  Knippschild, Lin, and Nagy}}]{Chu:2015nha}
\bibinfo{author}{\bibfnamefont{D.~Y.~J.} \bibnamefont{Chu}},
  \bibinfo{author}{\bibfnamefont{K.}~\bibnamefont{Jansen}},
  \bibinfo{author}{\bibfnamefont{B.}~\bibnamefont{Knippschild}},
  \bibinfo{author}{\bibfnamefont{C.~J.~D.} \bibnamefont{Lin}},
  \bibnamefont{and} \bibinfo{author}{\bibfnamefont{A.}~\bibnamefont{Nagy}},
  \bibinfo{journal}{Phys. Lett.} \textbf{\bibinfo{volume}{B744}},
  \bibinfo{pages}{146} (\bibinfo{year}{2015}), \eprint{1501.05440}.

\bibitem[{\citenamefont{Huang et~al.}(2016{\natexlab{a}})\citenamefont{Huang,
  Wan, Wang, Cai, and Zhang}}]{Huang:2016odd}
\bibinfo{author}{\bibfnamefont{F.~P.} \bibnamefont{Huang}},
  \bibinfo{author}{\bibfnamefont{Y.}~\bibnamefont{Wan}},
  \bibinfo{author}{\bibfnamefont{D.-G.} \bibnamefont{Wang}},
  \bibinfo{author}{\bibfnamefont{Y.-F.} \bibnamefont{Cai}}, \bibnamefont{and}
  \bibinfo{author}{\bibfnamefont{X.}~\bibnamefont{Zhang}},
  \bibinfo{journal}{Phys. Rev.} \textbf{\bibinfo{volume}{D94}},
  \bibinfo{pages}{041702} (\bibinfo{year}{2016}{\natexlab{a}}),
  \eprint{1601.01640}.

\bibitem[{\citenamefont{Quiros}(1999)}]{Quiros:1999jp}
\bibinfo{author}{\bibfnamefont{M.}~\bibnamefont{Quiros}}
  (\bibinfo{year}{1999}), \eprint{hep-ph/9901312}.

\bibitem[{\citenamefont{Dolan and Jackiw}(1974)}]{Dolan:1973qd}
\bibinfo{author}{\bibfnamefont{L.}~\bibnamefont{Dolan}} \bibnamefont{and}
  \bibinfo{author}{\bibfnamefont{R.}~\bibnamefont{Jackiw}},
  \bibinfo{journal}{Phys. Rev.} \textbf{\bibinfo{volume}{D9}},
  \bibinfo{pages}{3320} (\bibinfo{year}{1974}).

\bibitem[{\citenamefont{Spannowsky and Tamarit}(2017)}]{Spannowsky:2016ile}
\bibinfo{author}{\bibfnamefont{M.}~\bibnamefont{Spannowsky}} \bibnamefont{and}
  \bibinfo{author}{\bibfnamefont{C.}~\bibnamefont{Tamarit}},
  \bibinfo{journal}{Phys. Rev.} \textbf{\bibinfo{volume}{D95}},
  \bibinfo{pages}{015006} (\bibinfo{year}{2017}), \eprint{1611.05466}.

\bibitem[{\citenamefont{Gan et~al.}(2017)\citenamefont{Gan, Long, and
  Wang}}]{Gan:2017mcv}
\bibinfo{author}{\bibfnamefont{X.}~\bibnamefont{Gan}},
  \bibinfo{author}{\bibfnamefont{A.~J.} \bibnamefont{Long}}, \bibnamefont{and}
  \bibinfo{author}{\bibfnamefont{L.-T.} \bibnamefont{Wang}}
  (\bibinfo{year}{2017}), \eprint{1708.03061}.

\bibitem[{\citenamefont{Huang et~al.}(2016{\natexlab{b}})\citenamefont{Huang,
  Gu, Yin, Yu, and Zhang}}]{Huang:2015izx}
\bibinfo{author}{\bibfnamefont{F.~P.} \bibnamefont{Huang}},
  \bibinfo{author}{\bibfnamefont{P.-H.} \bibnamefont{Gu}},
  \bibinfo{author}{\bibfnamefont{P.-F.} \bibnamefont{Yin}},
  \bibinfo{author}{\bibfnamefont{Z.-H.} \bibnamefont{Yu}}, \bibnamefont{and}
  \bibinfo{author}{\bibfnamefont{X.}~\bibnamefont{Zhang}},
  \bibinfo{journal}{Phys. Rev.} \textbf{\bibinfo{volume}{D93}},
  \bibinfo{pages}{103515} (\bibinfo{year}{2016}{\natexlab{b}}),
  \eprint{1511.03969}.

\bibitem[{\citenamefont{Noble and Perelstein}(2008)}]{Noble:2007kk}
\bibinfo{author}{\bibfnamefont{A.}~\bibnamefont{Noble}} \bibnamefont{and}
  \bibinfo{author}{\bibfnamefont{M.}~\bibnamefont{Perelstein}},
  \bibinfo{journal}{Phys. Rev.} \textbf{\bibinfo{volume}{D78}},
  \bibinfo{pages}{063518} (\bibinfo{year}{2008}), \eprint{0711.3018}.

\bibitem[{\citenamefont{Katz and Perelstein}(2014)}]{Katz:2014bha}
\bibinfo{author}{\bibfnamefont{A.}~\bibnamefont{Katz}} \bibnamefont{and}
  \bibinfo{author}{\bibfnamefont{M.}~\bibnamefont{Perelstein}},
  \bibinfo{journal}{JHEP} \textbf{\bibinfo{volume}{07}}, \bibinfo{pages}{108}
  (\bibinfo{year}{2014}), \eprint{1401.1827}.

\bibitem[{\citenamefont{Curtin et~al.}(2014)\citenamefont{Curtin, Meade, and
  Yu}}]{Curtin:2014jma}
\bibinfo{author}{\bibfnamefont{D.}~\bibnamefont{Curtin}},
  \bibinfo{author}{\bibfnamefont{P.}~\bibnamefont{Meade}}, \bibnamefont{and}
  \bibinfo{author}{\bibfnamefont{C.-T.} \bibnamefont{Yu}},
  \bibinfo{journal}{JHEP} \textbf{\bibinfo{volume}{11}}, \bibinfo{pages}{127}
  (\bibinfo{year}{2014}), \eprint{1409.0005}.

\bibitem[{\citenamefont{Cai et~al.}(2017)\citenamefont{Cai, Sasaki, and
  Wang}}]{Cai:2017tmh}
\bibinfo{author}{\bibfnamefont{R.-G.} \bibnamefont{Cai}},
  \bibinfo{author}{\bibfnamefont{M.}~\bibnamefont{Sasaki}}, \bibnamefont{and}
  \bibinfo{author}{\bibfnamefont{S.-J.} \bibnamefont{Wang}},
  \bibinfo{journal}{JCAP} \textbf{\bibinfo{volume}{1708}}, \bibinfo{pages}{004}
  (\bibinfo{year}{2017}), \eprint{1707.03001}.

\bibitem[{\citenamefont{McCullough}(2014)}]{McCullough:2013rea}
\bibinfo{author}{\bibfnamefont{M.}~\bibnamefont{McCullough}},
  \bibinfo{journal}{Phys. Rev.} \textbf{\bibinfo{volume}{D90}},
  \bibinfo{pages}{015001} (\bibinfo{year}{2014}), \bibinfo{note}{[Erratum:
  Phys. Rev.D92,no.3,039903(2015)]}, \eprint{1312.3322}.

\bibitem[{\citenamefont{Englert and McCullough}(2013)}]{Englert:2013tya}
\bibinfo{author}{\bibfnamefont{C.}~\bibnamefont{Englert}} \bibnamefont{and}
  \bibinfo{author}{\bibfnamefont{M.}~\bibnamefont{McCullough}},
  \bibinfo{journal}{JHEP} \textbf{\bibinfo{volume}{07}}, \bibinfo{pages}{168}
  (\bibinfo{year}{2013}), \eprint{1303.1526}.

\bibitem[{\citenamefont{Baer et~al.}(2013)\citenamefont{Baer, Barklow, Fujii,
  Gao, Hoang, Kanemura, List, Logan, Nomerotski, Perelstein
  et~al.}}]{Baer:2013cma}
\bibinfo{author}{\bibfnamefont{H.}~\bibnamefont{Baer}},
  \bibinfo{author}{\bibfnamefont{T.}~\bibnamefont{Barklow}},
  \bibinfo{author}{\bibfnamefont{K.}~\bibnamefont{Fujii}},
  \bibinfo{author}{\bibfnamefont{Y.}~\bibnamefont{Gao}},
  \bibinfo{author}{\bibfnamefont{A.}~\bibnamefont{Hoang}},
  \bibinfo{author}{\bibfnamefont{S.}~\bibnamefont{Kanemura}},
  \bibinfo{author}{\bibfnamefont{J.}~\bibnamefont{List}},
  \bibinfo{author}{\bibfnamefont{H.~E.} \bibnamefont{Logan}},
  \bibinfo{author}{\bibfnamefont{A.}~\bibnamefont{Nomerotski}},
  \bibinfo{author}{\bibfnamefont{M.}~\bibnamefont{Perelstein}},
  \bibnamefont{et~al.} (\bibinfo{year}{2013}), \eprint{1306.6352}.

\bibitem[{\citenamefont{Azzi}(2014)}]{Azzi:2014jwa}
\bibinfo{author}{\bibfnamefont{P.}~\bibnamefont{Azzi}}, \bibinfo{journal}{Nuovo
  Cim.} \textbf{\bibinfo{volume}{C037}}, \bibinfo{pages}{11}
  (\bibinfo{year}{2014}).

\bibitem[{\citenamefont{Ruan}(2016)}]{Ruan:2014xxa}
\bibinfo{author}{\bibfnamefont{M.}~\bibnamefont{Ruan}}, \bibinfo{journal}{Nucl.
  Part. Phys. Proc.} \textbf{\bibinfo{volume}{273-275}}, \bibinfo{pages}{857}
  (\bibinfo{year}{2016}), \eprint{1411.5606}.

\bibitem[{\citenamefont{Gong et~al.}(2017)\citenamefont{Gong, Li, Xu, Yang, and
  Zhao}}]{Gong:2016jys}
\bibinfo{author}{\bibfnamefont{Y.}~\bibnamefont{Gong}},
  \bibinfo{author}{\bibfnamefont{Z.}~\bibnamefont{Li}},
  \bibinfo{author}{\bibfnamefont{X.}~\bibnamefont{Xu}},
  \bibinfo{author}{\bibfnamefont{L.~L.} \bibnamefont{Yang}}, \bibnamefont{and}
  \bibinfo{author}{\bibfnamefont{X.}~\bibnamefont{Zhao}},
  \bibinfo{journal}{Phys. Rev.} \textbf{\bibinfo{volume}{D95}},
  \bibinfo{pages}{093003} (\bibinfo{year}{2017}), \eprint{1609.03955}.

\bibitem[{\citenamefont{Sun et~al.}(2016)\citenamefont{Sun, Feng, Jia, and
  Sang}}]{Sun:2016bel}
\bibinfo{author}{\bibfnamefont{Q.-F.} \bibnamefont{Sun}},
  \bibinfo{author}{\bibfnamefont{F.}~\bibnamefont{Feng}},
  \bibinfo{author}{\bibfnamefont{Y.}~\bibnamefont{Jia}}, \bibnamefont{and}
  \bibinfo{author}{\bibfnamefont{W.-L.} \bibnamefont{Sang}}
  (\bibinfo{year}{2016}), \eprint{1609.03995}.

\bibitem[{\citenamefont{Craig et~al.}(2015)\citenamefont{Craig, Farina,
  McCullough, and Perelstein}}]{Craig:2014una}
\bibinfo{author}{\bibfnamefont{N.}~\bibnamefont{Craig}},
  \bibinfo{author}{\bibfnamefont{M.}~\bibnamefont{Farina}},
  \bibinfo{author}{\bibfnamefont{M.}~\bibnamefont{McCullough}},
  \bibnamefont{and}
  \bibinfo{author}{\bibfnamefont{M.}~\bibnamefont{Perelstein}},
  \bibinfo{journal}{JHEP} \textbf{\bibinfo{volume}{03}}, \bibinfo{pages}{146}
  (\bibinfo{year}{2015}), \eprint{1411.0676}.

\bibitem[{\citenamefont{Ge et~al.}(2016)\citenamefont{Ge, He, and
  Xiao}}]{Ge:2016zro}
\bibinfo{author}{\bibfnamefont{S.-F.} \bibnamefont{Ge}},
  \bibinfo{author}{\bibfnamefont{H.-J.} \bibnamefont{He}}, \bibnamefont{and}
  \bibinfo{author}{\bibfnamefont{R.-Q.} \bibnamefont{Xiao}},
  \bibinfo{journal}{JHEP} \textbf{\bibinfo{volume}{10}}, \bibinfo{pages}{007}
  (\bibinfo{year}{2016}), \eprint{1603.03385}.

\bibitem[{\citenamefont{Georgi and Machacek}(1985)}]{Georgi:1985nv}
\bibinfo{author}{\bibfnamefont{H.}~\bibnamefont{Georgi}} \bibnamefont{and}
  \bibinfo{author}{\bibfnamefont{M.}~\bibnamefont{Machacek}},
  \bibinfo{journal}{Nucl. Phys.} \textbf{\bibinfo{volume}{B262}},
  \bibinfo{pages}{463} (\bibinfo{year}{1985}).

\bibitem[{\citenamefont{Cao et~al.}(2016)\citenamefont{Cao, Liu, Xie, Yan, and
  Zhang}}]{Cao:2015scs}
\bibinfo{author}{\bibfnamefont{Q.-H.} \bibnamefont{Cao}},
  \bibinfo{author}{\bibfnamefont{Y.}~\bibnamefont{Liu}},
  \bibinfo{author}{\bibfnamefont{K.-P.} \bibnamefont{Xie}},
  \bibinfo{author}{\bibfnamefont{B.}~\bibnamefont{Yan}}, \bibnamefont{and}
  \bibinfo{author}{\bibfnamefont{D.-M.} \bibnamefont{Zhang}},
  \bibinfo{journal}{Phys. Rev.} \textbf{\bibinfo{volume}{D93}},
  \bibinfo{pages}{075030} (\bibinfo{year}{2016}), \eprint{1512.08441}.

\bibitem[{\citenamefont{Cao and Zhang}(2016)}]{Cao:2016uur}
\bibinfo{author}{\bibfnamefont{Q.-H.} \bibnamefont{Cao}} \bibnamefont{and}
  \bibinfo{author}{\bibfnamefont{D.-M.} \bibnamefont{Zhang}}
  (\bibinfo{year}{2016}), \eprint{1611.09337}.

\bibitem[{\citenamefont{Huang and Zhang}(2017)}]{Huang:2017laj}
\bibinfo{author}{\bibfnamefont{F.~P.} \bibnamefont{Huang}} \bibnamefont{and}
  \bibinfo{author}{\bibfnamefont{X.}~\bibnamefont{Zhang}}
  (\bibinfo{year}{2017}), \eprint{1701.04338}.

\bibitem[{\citenamefont{Khandker et~al.}(2012)\citenamefont{Khandker, Li, and
  Skiba}}]{Khandker:2012zu}
\bibinfo{author}{\bibfnamefont{Z.~U.} \bibnamefont{Khandker}},
  \bibinfo{author}{\bibfnamefont{D.}~\bibnamefont{Li}}, \bibnamefont{and}
  \bibinfo{author}{\bibfnamefont{W.}~\bibnamefont{Skiba}},
  \bibinfo{journal}{Phys. Rev.} \textbf{\bibinfo{volume}{D86}},
  \bibinfo{pages}{015006} (\bibinfo{year}{2012}), \eprint{1201.4383}.

\bibitem[{\citenamefont{Khan}(2016)}]{Khan:2016sxm}
\bibinfo{author}{\bibfnamefont{N.}~\bibnamefont{Khan}} (\bibinfo{year}{2016}),
  \eprint{1610.03178}.

\bibitem[{\citenamefont{Jenkins et~al.}(2013)\citenamefont{Jenkins, Manohar,
  and Trott}}]{Jenkins:2013zja}
\bibinfo{author}{\bibfnamefont{E.~E.} \bibnamefont{Jenkins}},
  \bibinfo{author}{\bibfnamefont{A.~V.} \bibnamefont{Manohar}},
  \bibnamefont{and} \bibinfo{author}{\bibfnamefont{M.}~\bibnamefont{Trott}},
  \bibinfo{journal}{JHEP} \textbf{\bibinfo{volume}{10}}, \bibinfo{pages}{087}
  (\bibinfo{year}{2013}), \eprint{1308.2627}.

\bibitem[{\citenamefont{Jenkins et~al.}(2014)\citenamefont{Jenkins, Manohar,
  and Trott}}]{Jenkins:2013wua}
\bibinfo{author}{\bibfnamefont{E.~E.} \bibnamefont{Jenkins}},
  \bibinfo{author}{\bibfnamefont{A.~V.} \bibnamefont{Manohar}},
  \bibnamefont{and} \bibinfo{author}{\bibfnamefont{M.}~\bibnamefont{Trott}},
  \bibinfo{journal}{JHEP} \textbf{\bibinfo{volume}{01}}, \bibinfo{pages}{035}
  (\bibinfo{year}{2014}), \eprint{1310.4838}.

\bibitem[{\citenamefont{Baak et~al.}(2014)\citenamefont{Baak, Cuth, Haller,
  Hoecker, Kogler, Monig, Schott, and Stelzer}}]{Baak:2014ora}
\bibinfo{author}{\bibfnamefont{M.}~\bibnamefont{Baak}},
  \bibinfo{author}{\bibfnamefont{J.}~\bibnamefont{Cuth}},
  \bibinfo{author}{\bibfnamefont{J.}~\bibnamefont{Haller}},
  \bibinfo{author}{\bibfnamefont{A.}~\bibnamefont{Hoecker}},
  \bibinfo{author}{\bibfnamefont{R.}~\bibnamefont{Kogler}},
  \bibinfo{author}{\bibfnamefont{K.}~\bibnamefont{Monig}},
  \bibinfo{author}{\bibfnamefont{M.}~\bibnamefont{Schott}}, \bibnamefont{and}
  \bibinfo{author}{\bibfnamefont{J.}~\bibnamefont{Stelzer}}
  (\bibinfo{collaboration}{Gfitter Group}), \bibinfo{journal}{Eur. Phys. J.}
  \textbf{\bibinfo{volume}{C74}}, \bibinfo{pages}{3046} (\bibinfo{year}{2014}),
  \eprint{1407.3792}.

\bibitem[{\citenamefont{Andersen et~al.}(2013)}]{Heinemeyer:2013tqa}
\bibinfo{author}{\bibfnamefont{J.~R.} \bibnamefont{Andersen}}
  \bibnamefont{et~al.} (\bibinfo{collaboration}{LHC Higgs Cross Section Working
  Group}) (\bibinfo{year}{2013}), \eprint{1307.1347}.

\bibitem[{\citenamefont{Schael et~al.}(2006)}]{ALEPH:2005ab}
\bibinfo{author}{\bibfnamefont{S.}~\bibnamefont{Schael}} \bibnamefont{et~al.}
  (\bibinfo{collaboration}{SLD Electroweak Group, DELPHI, ALEPH, SLD, SLD Heavy
  Flavour Group, OPAL, LEP Electroweak Working Group, L3}),
  \bibinfo{journal}{Phys. Rept.} \textbf{\bibinfo{volume}{427}},
  \bibinfo{pages}{257} (\bibinfo{year}{2006}), \eprint{hep-ex/0509008}.

\bibitem[{\citenamefont{Group}(2012)}]{Group:2012gb}
\bibinfo{author}{\bibfnamefont{T.~E.~W.} \bibnamefont{Group}}
  (\bibinfo{collaboration}{CDF, D0}) (\bibinfo{year}{2012}),
  \eprint{1204.0042}.

\bibitem[{\citenamefont{Aad et~al.}(2016)}]{Khachatryan:2016vau}
\bibinfo{author}{\bibfnamefont{G.}~\bibnamefont{Aad}} \bibnamefont{et~al.}
  (\bibinfo{collaboration}{ATLAS, CMS}), \bibinfo{journal}{JHEP}
  \textbf{\bibinfo{volume}{08}}, \bibinfo{pages}{045} (\bibinfo{year}{2016}),
  \eprint{1606.02266}.

\bibitem[{\citenamefont{Di~Luzio et~al.}(2015)\citenamefont{Di~Luzio, Grober,
  Kamenik, and Nardecchia}}]{DiLuzio:2015oha}
\bibinfo{author}{\bibfnamefont{L.}~\bibnamefont{Di~Luzio}},
  \bibinfo{author}{\bibfnamefont{R.}~\bibnamefont{Grober}},
  \bibinfo{author}{\bibfnamefont{J.~F.} \bibnamefont{Kamenik}},
  \bibnamefont{and}
  \bibinfo{author}{\bibfnamefont{M.}~\bibnamefont{Nardecchia}},
  \bibinfo{journal}{JHEP} \textbf{\bibinfo{volume}{07}}, \bibinfo{pages}{074}
  (\bibinfo{year}{2015}), \eprint{1504.00359}.

\bibitem[{\citenamefont{Ma}(2006)}]{Ma:2006km}
\bibinfo{author}{\bibfnamefont{E.}~\bibnamefont{Ma}}, \bibinfo{journal}{Phys.
  Rev.} \textbf{\bibinfo{volume}{D73}}, \bibinfo{pages}{077301}
  (\bibinfo{year}{2006}), \eprint{hep-ph/0601225}.

\bibitem[{\citenamefont{Barbieri et~al.}(2006)\citenamefont{Barbieri, Hall, and
  Rychkov}}]{Barbieri:2006dq}
\bibinfo{author}{\bibfnamefont{R.}~\bibnamefont{Barbieri}},
  \bibinfo{author}{\bibfnamefont{L.~J.} \bibnamefont{Hall}}, \bibnamefont{and}
  \bibinfo{author}{\bibfnamefont{V.~S.} \bibnamefont{Rychkov}},
  \bibinfo{journal}{Phys. Rev.} \textbf{\bibinfo{volume}{D74}},
  \bibinfo{pages}{015007} (\bibinfo{year}{2006}), \eprint{hep-ph/0603188}.

\bibitem[{\citenamefont{Cao et~al.}(2007)\citenamefont{Cao, Ma, and
  Rajasekaran}}]{Cao:2007rm}
\bibinfo{author}{\bibfnamefont{Q.-H.} \bibnamefont{Cao}},
  \bibinfo{author}{\bibfnamefont{E.}~\bibnamefont{Ma}}, \bibnamefont{and}
  \bibinfo{author}{\bibfnamefont{G.}~\bibnamefont{Rajasekaran}},
  \bibinfo{journal}{Phys. Rev.} \textbf{\bibinfo{volume}{D76}},
  \bibinfo{pages}{095011} (\bibinfo{year}{2007}), \eprint{0708.2939}.

\bibitem[{\citenamefont{Barroso et~al.}(2013)\citenamefont{Barroso, Ferreira,
  Ivanov, and Santos}}]{Barroso:2013awa}
\bibinfo{author}{\bibfnamefont{A.}~\bibnamefont{Barroso}},
  \bibinfo{author}{\bibfnamefont{P.~M.} \bibnamefont{Ferreira}},
  \bibinfo{author}{\bibfnamefont{I.~P.} \bibnamefont{Ivanov}},
  \bibnamefont{and} \bibinfo{author}{\bibfnamefont{R.}~\bibnamefont{Santos}},
  \bibinfo{journal}{JHEP} \textbf{\bibinfo{volume}{06}}, \bibinfo{pages}{045}
  (\bibinfo{year}{2013}), \eprint{1303.5098}.

\bibitem[{\citenamefont{Chen et~al.}(2017)\citenamefont{Chen, Kozaczuk, and
  Lewis}}]{Chen:2017qcz}
\bibinfo{author}{\bibfnamefont{C.-Y.} \bibnamefont{Chen}},
  \bibinfo{author}{\bibfnamefont{J.}~\bibnamefont{Kozaczuk}}, \bibnamefont{and}
  \bibinfo{author}{\bibfnamefont{I.~M.} \bibnamefont{Lewis}},
  \bibinfo{journal}{JHEP} \textbf{\bibinfo{volume}{08}}, \bibinfo{pages}{096}
  (\bibinfo{year}{2017}), \eprint{1704.05844}.

\bibitem[{ATL(2015)}]{ATLAS-CONF-2015-044}
\bibinfo{type}{Tech. Rep.} \bibinfo{number}{ATLAS-CONF-2015-044},
  \bibinfo{institution}{CERN}, \bibinfo{address}{Geneva}
  (\bibinfo{year}{2015}), \urlprefix\url{http://cds.cern.ch/record/2052552}.

\bibitem[{ATL(2017)}]{ATLAS-CONF-2017-045}
\bibinfo{type}{Tech. Rep.} \bibinfo{number}{ATLAS-CONF-2017-045},
  \bibinfo{institution}{CERN}, \bibinfo{address}{Geneva}
  (\bibinfo{year}{2017}), \urlprefix\url{http://cds.cern.ch/record/2273852}.

\bibitem[{\citenamefont{Damgaard et~al.}(2016)\citenamefont{Damgaard, Haarr,
  O'Connell, and Tranberg}}]{Damgaard:2015con}
\bibinfo{author}{\bibfnamefont{P.~H.} \bibnamefont{Damgaard}},
  \bibinfo{author}{\bibfnamefont{A.}~\bibnamefont{Haarr}},
  \bibinfo{author}{\bibfnamefont{D.}~\bibnamefont{O'Connell}},
  \bibnamefont{and} \bibinfo{author}{\bibfnamefont{A.}~\bibnamefont{Tranberg}},
  \bibinfo{journal}{JHEP} \textbf{\bibinfo{volume}{02}}, \bibinfo{pages}{107}
  (\bibinfo{year}{2016}), \eprint{1512.01963}.

\bibitem[{\citenamefont{Georgi}(1993)}]{Georgi:1992dw}
\bibinfo{author}{\bibfnamefont{H.}~\bibnamefont{Georgi}},
  \bibinfo{journal}{Phys. Lett.} \textbf{\bibinfo{volume}{B298}},
  \bibinfo{pages}{187} (\bibinfo{year}{1993}), \eprint{hep-ph/9207278}.

\end{thebibliography}
\end{document}